\def\nk{n_{\rm b}}
\def\acap{\\ \nonumber \\}
\def\beg{\nonumber &}
\def\kl{\mathbf{\hat{k}}\mathbf\cdot\mathbf{\hat{l}}}
\def\km{\mathbf{\hat{k}}\mathbf\cdot\mathbf{\hat{m}}}
\def\kh{\mathbf{\hat{k}}\mathbf\cdot\mathbf{\hat{h}}}
\def\Pb{P_{\rm b}}
\def\rfr#1{Equation\,(\ref{#1})}
\def\rfrs#1#2{Equations\,(\ref{#1})--(\ref{#2})}
\def\Rfr#1{Equation\,(\ref{#1})}
\def\Rfrs#1#2{Equations\,(\ref{#1})--(\ref{#2})}
\def\derp#1#2{\rp{\partial{#1}}{\partial{#2}}}
\def\dert#1#2{\frac{{{\textrm{d}}}{#1}}{{{\textrm{d}}}{#2}}}
\def\virg#1{``#1"}
\def\eqi{\begin{equation}}
\def\eqf{\end{equation}}
\def\rp#1#2{\frac{#1}{#2}}
\def\lb#1{\label{#1}}
\def\bds#1{\mathbf{#1}}
\def\ton#1{\left(#1\right)}
\def\qua#1{\left[#1\right]}
\def\grf#1{\left\{#1\right\}}
\def\ang#1{\left\langle #1\right\rangle}
\RequirePackage[2020-02-02]{latexrelease}
\documentclass{aastex631}
\usepackage{morefloats}
\usepackage[title]{appendix}
\usepackage{textcomp}
\usepackage{booktabs}
\usepackage{multirow}
\usepackage{rotating,tabularx}
\usepackage{float}
\usepackage{enumerate}
\usepackage{rotating}
\usepackage[polutonikogreek,english]{babel}
\usepackage{amsmath,starfont,textgreek,w-greek,wasysym}
\usepackage[flushleft]{threeparttable}
\usepackage{amsthm}
\usepackage{amscd}
\usepackage[mathlines]{lineno}
\usepackage{amssymb,dsfont}
\usepackage{graphicx,epsfig}
\usepackage{txfonts}
\usepackage{xr-hyper}
\usepackage{hyperref}
\allowdisplaybreaks

\makeatletter
 \DeclareRobustCommand\ref{%
    \@ifstar\@refstar\T@ref
  }%
  \DeclareRobustCommand\pageref{%
    \@ifstar\@pagerefstar\T@pageref
  }%
 \makeatother

\begin{document}

\title{The post-Newtonian motion around an oblate spheroid: the mixed orbital effects due to the Newtonian oblateness and the post-Newtonian mass monopole accelerations}

\shortauthors{L. Iorio}

\author[0000-0003-4949-2694]{Lorenzo Iorio}
\affiliation{Ministero dell' Istruzione e del Merito. Viale Unit\`{a} di Italia 68, I-70125, Bari (BA),
Italy}

\email{lorenzo.iorio@libero.it}

\begin{abstract}
When a test particle moves about an oblate spheroid, it is acted upon, among other things, by two standard perturbing accelerations. One, of Newtonian origin, is due to the quadrupole mass moment $J_2$ of the orbited body. The other one, \textcolor{black}{of order} $\mathcal{O}\left(1/c^2\right)$, is caused by the static, post-Newtonian field arising solely from the mass of the central object. Both of them concur to induce \textrm{indirect}, \textrm{mixed} orbital effects \textcolor{black}{of order} $\mathcal{O}\left(J_2/c^2\right)$. They are of the same order of magnitude of the \textrm{direct} ones induced by the post-Newtonian acceleration
%\textcolor{black}{of order} $\mathcal{O}\ton{J_2/c^2}$
arising in presence of an oblate source, not treated here. We calculate these less known features of motion in their full generality in terms of the osculating Keplerian orbital elements. Subtleties pertaining the correct calculation of their mixed net \textrm{precessions} per orbit to the full order of $\mathcal{O}\left(J_2/c^2\right)$ are elucidated. The obtained results hold for arbitrary orbital geometries and for any orientation of the body's spin axis $\mathbf{\hat{k}}$ in space. The method presented is completely general, and can be extended to any pair of post-Keplerian accelerations entering the equations of motion of the satellite, irrespectively of their physical nature.
\end{abstract}

%{
%\textrm{Unified Astronomy Thesaurus concepts}:\,Exoplanets\,(498); General relativity\,(641)
%}
%\centerline
%{PACS: 04.20.-q; 04.20.Cv; 04.80.-y; 04.80.Cc; 91.10.Sp}

\section{Introduction}
To the first post-Newtonian (1pN) order, the quadrupole mass moment $J_2$ of an oblate spheroid that is rigidly rotating causes an acceleration of the order $\mathcal{O}\ton{J_2/c^2}$,
%\citep{1988CeMec..42...81S,Sof89,1991ercm.book.....B,1992CeMDA..53..293H,2014PhRvD..89d4043W},
where $c$ is the speed of light in vacuum, which \textrm{directly} induces long-term orbital variations affecting the motion of a test particle. They have been treated to various levels of completeness  elsewhere \citep{1988CeMec..42...81S,Sof89,1991ercm.book.....B,1992CeMDA..53..293H,2014PhRvD..89d4043W,2015IJMPD..2450067I}\textcolor{black}{; the most general calculation, valid for any orbital geometry and arbitrary orientation of the primary's spin axis can be found in \citet{2023arXiv231002834I}.}

Nonetheless, further \textrm{indirect} orbital features of motion \textcolor{black}{of order} $\mathcal{O}\ton{J_2/c^2}$ arise due to the interplay of two well known post-Keplerian (pK) accelerations: the Newtonian one induced by $J_2$, causing the orbital plane of an Earth's satellite to secularly precess \citep{1958RSPSA.247...49K,Capde05,2014grav.book.....P}, and the 1pN \virg{gravitoelectric} term due to the mass monopole of the body \citep{Sof89,2014grav.book.....P,SoffelHan19},  responsible of the formerly anomalous perihelion precession of Mercury \citep{LeVer1859} of $42.98$ arcseconds per century (arcsec cty$^{-1}$) \citep{1986Natur.320...39N}. \textcolor{black}{Its explanation by \citet{Ein15} was the first empirical success of his newborn General Theory of Relativity (GTR). For a recent review, see, e.g., \citet{2016Univ....2...23D}, and references therein.} Calculations for the \virg{\textrm{mixed}} effects \textcolor{black}{resulting from the interplay of the aforementioned pK accelerations}, performed mainly recurring to some simplifying assumptions and to various computational schemes, can be found in the literature \citep{1990CeMDA..47..205H,1992CeMDA..53..293H,2014PhRvD..89d4043W,2015IJMPD..2450067I}.

The task of the present work is to calculate them in their full generality, elucidating certain subtleties occurring when one moves from the orbital \textrm{shifts per revolution} to the averaged orbital \textrm{precessions}.

The paper is organized as follows. In Section\,\ref{shifts}, a general overview of the calculation of the mixed effects due to a pair of arbitrary disturbing accelerations is first presented (Section\,\ref{over}). Then, it is applied to the aforementioned pK perturbations, and explicit expressions for the \textrm{mixed} net \textrm{shifts per orbit} \textcolor{black}{of order} $\mathcal{O}\ton{J_2/c^2}$ of  all the Keplerian orbital elements, valid for any orbital configurations and arbitrary orientations of the body's symmetry axis in space, are displayed (Section\,\ref{j2c2}). Section\,\ref{rates} is devoted to the calculation of the \textrm{total} mixed averaged \textrm{rates of change} \textcolor{black}{of order} $\mathcal{O}\ton{J_2/c^2}$  of the orbital elements by elucidating that it is not enough to simply take the ratios of the averaged variations obtained in Section\,\ref{shifts} to the Keplerian orbital period. Also in this case, explicit expressions of general validity are obtained. In Section\,\ref{orbi}, the results obtained in Sections\,\ref{shifts}\,to\,\ref{rates} are specialized to two particular  configurations: equatorial (Section\,\ref{equa}) and polar (Section\,\ref{pola}) orbits. Section\,\ref{fine} summarizes our results and offers concluding remarks.
\section{The mixed net shifts per orbit}\lb{shifts}
\subsection{General calculational overview}\lb{over}
Let us assume that a perturbing acceleration
\eqi
\bds A = {\bds A}^\mathrm{\ton{1}} + {\bds A}^\mathrm{\ton{2}},
\eqf
made of the sum of two pK accelerations ${\bds A}^\mathrm{\ton{1}}$ and ${\bds A}^\mathrm{\ton{2}}$ of arbitrary origin,
enters the equations of motion of a test particle orbiting a central body in addition to the dominant Newtonian monopole.
%For example, ${\bds A}^\mathrm{\ton{1}}$ and ${\bds A}^\mathrm{\ton{2}}$  may be ${\bds A}^{\ton{J_2}}$ induced by the quadrupole mass moment $J_2$ of the %primary, and the 1pN \virg{gravitoelectric} acceleration ${\bds A}^{\ton{\mathrm{1pN}}}$ due to the mass monopole of the body, as in the present case.
 The osculating Keplerian orbital elements $\grf{\kappa}$ of the satellite, i.e. the semimajor axis $a$, the eccentricity $e$, the inclination $I$ of the orbital plane to the reference plane $\grf{x,\,y}$, the longitude of the ascending node $\Omega$, the argument of pericenter $\omega$, and the mean anomaly at epoch $\eta$ \citep{1994ApJ...427..951K,2011rcms.book.....K}, undergo long-term variations\textcolor{black}{; for non-osculating orbital elements in pN dynamics, see \citet{2022AdSpR..69..538G}. These long-term variations of any one of the Keplerian
orbital elements $\grf{\kappa}$ can be calculated as averages} over one orbital period $\Pb$,  that are caused not only by each of the two pK accelerations ${\bds A}^\mathrm{\ton{1}}$ and ${\bds A}^\mathrm{\ton{2}}$ individually as if the other were not present, but also by the simultaneous action of both of them giving rise to indirect, mixed effects\textcolor{black}{. It means}
\eqi
\textcolor{black}{
\Delta\kappa = \int_{f_0}^{f_0 + 2\pi}\,\ton{\dert{\kappa}{f}}^{\mathrm{\ton{1}}}_\mathrm{K}\,\mathrm{d}f + \int_{f_0}^{f_0 + 2\pi}\,\ton{\dert{\kappa}{f}}^{\mathrm{\ton{2}}}_\mathrm{K}\,\mathrm{d}f + \int_{f_0}^{f_0 + 2\pi}\,\ton{\dert{\kappa}{f}}^{\mathrm{\ton{1-2}}}_\mathrm{mix}\,\mathrm{d}f,
}\lb{referee2}
\eqf
\textcolor{black}{
where $f$ is the true anomaly and $f_0$ is the anomaly at some arbitrary moment of time
$t_0$ assumed as initial instant. Here and in the following, the angular brackets $\ang{\cdots}$
denoting the average over $\Pb$ will be neglected in order to make the overall notation less cumbersome. \textcolor{black}{As it will become clearer below, it is worthwhile noticing that the average is taken over the anomalistic period, i.e., the time elapsed between two successive crossings of the moving pericenter due to the pK acceleration(s).}
In our work we are interested in the mixed average term, that is
}
\eqi
\textcolor{black}
{
\Delta\kappa^{\mathrm{\ton{1-2}}}_\mathrm{mix} :=\int_{f_0}^{f_0 + 2\pi}\,\ton{\dert{\kappa}{f}}^{\mathrm{\ton{1-2}}}_\mathrm{mix}\,\mathrm{d}f,\lb{eq1}
}
\eqf
where
%\footnote{To the first order in both the pK accelerations, $\mathrm{d}\kappa/\mathrm{d}f$ can be expanded as $\mathrm{d}\kappa/\mathrm{d}f = %\grf{\mathrm{d}\kappa/\mathrm{d}f}^{\ton{1}}_\mathrm{K} + \grf{\mathrm{d}\kappa/\mathrm{d}f}^{\ton{2}}_\mathrm{K} + %\ton{\mathrm{d}\kappa/\mathrm{d}f}^{\mathrm{\ton{1-2}}}_\mathrm{mix}$, where the last term is given by \rfr{eq2}.}
\begin{align}
\ton{\dert{\kappa}{f}}^{\mathrm{\ton{1-2}}}_\mathrm{mix}\nonumber \lb{eq2}&:=\sum_{j=a,\,e,\,I,\,\Omega,\,\omega}\grf{\derp{\ton{\mathrm{d}\kappa/\mathrm{d}f}^\mathrm{\ton{1}}}{\kappa_j}}_\mathrm{K}\Delta\kappa_j^\mathrm{\ton{2}}\ton{f} + \acap
\nonumber & +\grf{\ton{\dert{\kappa}{f}}^\mathrm{\ton{1}}\rp{r^2}{\mu\,e}\qua{-\cos f\,A_R^\mathrm{\ton{2}} +\ton{1 + \rp{r}{p}}\,\sin f\,A_T^\mathrm{\ton{2}} }}_\mathrm{K}  + \\ \nonumber \\
\nonumber & \textcolor{black}{
+ \sum_{j=a,\,e,\,I,\,\Omega,\,\omega}\grf{\derp{\ton{\mathrm{d}\kappa/\mathrm{d}f}^\mathrm{\ton{2}}}{\kappa_j}}_\mathrm{K}\Delta\kappa_j^\mathrm{\ton{1}}\ton{f} +
} \acap
& \textcolor{black}{
+\grf{\ton{\dert{\kappa}{f}}^\mathrm{\ton{2}}\rp{r^2}{\mu\,e}\qua{-\cos f\,A_R^\mathrm{\ton{1}} +\ton{1 + \rp{r}{p}}\,\sin f\,A_T^\mathrm{\ton{1}} }}_\mathrm{K}.
}
\end{align}
\textcolor{black}{
It should be noted that $\partial\ton{\mathrm{d}\kappa/\mathrm{d}f}/\partial\eta = 0$ for all the Keplerian orbital elements and for any of the pK accelerations considered here; thus, the summations in \rfr{eq2} do not run over $\eta$ as well.
}
%In \rfr{eq2}, the term denoted as $\ton{1} \leftrightarrows\,\ton{2}$ is the same as the one explicitly displayed before it with ${\bds %A}^{\mathrm{\ton{1}}}$ replaced by ${\bds A}^{\mathrm{\ton{2}}}$, and vice versa.
In \textcolor{black}{\rfr{eq2}},
%$f$ is the true anomaly, $f_0$ is the true anomaly at some arbitrary moment of time $t_0$ assumed as initial instant,
\eqi
\mu:= G\,M
\eqf
is the primary's gravitational parameter given by the product of the Newtonian constant of gravitation $G$ by its mass $M$,
\eqi
p:= a\,\ton{1 - e^2}
\eqf
is the orbit's semilatus rectum, and $A_R,\,A_T$ are the radial and transverse components of the pK acceleration at hand, respectively. Furthermore, the subscript $\virg{_\mathrm{K}}$ means that the quantities in curly brackets to which it is appended are to be evaluated onto the Keplerian ellipse
\eqi
r =\rp{p}{1 + e\,\cos f},\lb{rkep}
\eqf
assumed as unperturbed, reference trajectory. Finally, the derivative $\mathrm{d}\kappa/\mathrm{d}f$ of any one of the orbital elements $\kappa$ with respect to the true anomaly $f$ is given by
\eqi
\dert\kappa f =\dert\kappa t\,\textcolor{black}{\ton{\dert t f}_\mathrm{K}},\lb{dkdf}
\eqf
where
\eqi
\textcolor{black}{\ton{\dert t f}_\mathrm{K} :=\rp{r^2}{\sqrt{\mu\,p}}}\lb{dtdf}
\eqf
is the Keplerian expression for the reciprocal of the time derivative of the true anomaly, and $\mathrm{d}\kappa/\mathrm{d}t$ is given by
the right-hand-side of the corresponding Gaussian equation for its variation. The equations for the variations of the Keplerian osculating elements in the Euler-Gauss form \citep{Sof89,1991ercm.book.....B,2011rcms.book.....K,SoffelHan19} are
\begin{align}
\dert{a}{t} \lb{dadt} &= \rp{2}{\nk^{\ton{\mathrm{K}}}\,\sqrt{1-e^2}}\,\qua{e\,A_R\,\sin f + \ton{\rp{p}{r}}\,A_T},\\ \nonumber \\
\dert e t \lb{dedt} & = \rp{\sqrt{1-e^2}}{\nk^{\ton{\mathrm{K}}}\,a}\,\grf{A_R\,\sin f + A_T\,\qua{\cos f + \rp{1}{e}\,\ton{1-\rp{r}{a}} }}, \\ \nonumber\\
\dert I t & = \rp{1}{\nk^{\ton{\mathrm{K}}}\,a\,\sqrt{1-e^2}}\,A_N\,\ton{\rp{r}{a}}\,\cos u, \\ \nonumber\\
\dert \Omega t \lb{dOdt}& = \rp{1}{\nk^{\ton{\mathrm{K}}}\,a\,\sin I\,\sqrt{1-e^2}}\,A_N\,\ton{\rp{r}{a}}\,\sin u, \\ \nonumber\\
\dert \omega t \lb{dodt} & = \rp{\sqrt{1-e^2}}{\nk^{\ton{\mathrm{K}}}\,a\,e}\,\qua{-A_R\,\cos f + A_T\,\ton{1 + \rp{r}{p}}\,\sin f}- \cos I\,\dert\Omega t, \\ \nonumber \\
\dert\eta{t} \lb{detadt} &= -\rp{2}{\nk^{\ton{\mathrm{K}}}\,a}\,A_R\,\ton{\rp{r}{a}} - \rp{\ton{1-e^2}}{\nk\textcolor{black}{^{\ton{\mathrm{K}}}}\,a\,e}\,\qua{-A_R\,\cos f +A_T\,\ton{1+\rp{r}{p}}\,\sin f}.
\end{align}
In \rfrs{dadt}{detadt},
\eqi
\nk^{\ton{\mathrm{K}}}= \rp{2\,\pi}{\Pb^{\ton{\mathrm{K}}}}=\sqrt{\rp{\mu}{a^3}}\lb{nkep}
\eqf
is the Keplerian mean motion, which is proportional to the reciprocal of the Keplerian orbital period $\Pb^{\ton{\mathrm{K}}}$,
\eqi
u:=\omega + f
\eqf
is the argument of latitude,
and $A_N$ is the normal component of the pK acceleration at hand.
In \rfr{eq2}, the instantaneous variations of the Keplerian orbital elements $\Delta\kappa^{\mathrm{\ton{pK}}}_j\ton{f}$ are present; they can be calculated for any of them as
\eqi
\Delta\kappa^{\mathrm{\ton{pK}}}\ton{f} = \int_{f_0}^f\ton{\dert{\kappa}{f^{'}}}^{\mathrm{\ton{pK}}}\,\mathrm{d}f^{'},\lb{Dkappa}
\eqf
where the derivative, taken from \rfrs{dadt}{detadt}, has to be evaluated onto the Keplerian ellipse of \rfr{rkep} for a given pK acceleration\textcolor{black}{; in this case, the subscript $\virg{_\mathrm{K}}$ is omitted to avoid making the notation too heavy}.

\textcolor{black}{Some explanatory remarks about the structure of \rfr{eq2} are, now, in order.
The first two terms of \rfr{referee2}, and the second and the fourth terms in \rfr{eq2} come from
\begin{align}
\dert\kappa f \nonumber &= \dert\kappa t\,\dert t f = \dert\kappa t\,\ton{\dert t f}_\mathrm{K}\,\rp{1}{1-\dert\omega f- \cos I\,\dert\Omega f}\simeq \ton{\dert \kappa f}_\mathrm{K}\,\ton{1 + \dert\omega f + \cos I\,\dert\Omega f} = \\ \nonumber \\
&=\ton{\dert \kappa f}_\mathrm{K}\,\grf{1 + \rp{r^2}{\mu\,e}\,\qua{-\cos f\,A_R + \ton{1 + \rp{r}{p}}\,\sin f\,A_T}}.\lb{ugh}
\end{align}
In obtaining \rfr{ugh},
the pK expression \citep{1991ercm.book.....B,2003ASSL..293.....B,2014grav.book.....P}
\eqi
\dert f t = \ton{\dert f t}_\mathrm{K} -\dert\omega t -\cos I\,\dert\Omega t = \rp{\sqrt{\mu\,p}}{r^2} -\dert\omega t -\cos I\,\dert\Omega t, \lb{brumby}
\eqf
\rfr{dkdf}, and \rfrs{dOdt}{dodt} were used. \Rfr{brumby} accounts for the fact that, in general,  both the nodal and apsidal lines vary instantaneously during an orbital revolution because of pK perturbing accelerations. It does matter since the fast variable of integration is the true anomaly $f$; as remarked before, the averaging time interval, is the anomalistic period.
}
\textcolor{black}{
Going into more detail, t}he superscripts $\virg{^\mathrm{\ton{1}}}$ and $\virg{^\mathrm{\ton{2}}}$ in \rfr{eq2} mean that the associated quantities have to be calculated with the pK acceleration ${\bds A}^\mathrm{\ton{1}}$ and ${\bds A}^\mathrm{\ton{2}}$, respectively. \textcolor{black}{Thus, $\partial\ton{\mathrm{d}\kappa/\mathrm{d}f}^{\ton{1}}/\partial\kappa_j$ and $\partial\ton{\mathrm{d}\kappa/\mathrm{d}f}^{\ton{2}}/\partial\kappa_j$
are to be meant as the partial derivatives of \rfr{dkdf}
%, obtained from the right-hand-sides of \rfrs{dadt}{detadt} and \rfr{dtdf},
with respect to $\kappa_j,\,j = a,\,e,\,I,\,\Omega,\,\omega$ calculated  with the accelerations ${\bds A}^\mathrm{\ton{1}}$ and ${\bds A}^\mathrm{\ton{2}}$, respectively, onto the Keplerian ellipse.
The first and the third terms of \rfr{eq2}, arising from such partial derivatives, occur because, actually, the orbital elements do not stay constant during an  orbital revolution;
%, as when one uses an unperturbed Keplerian ellipse in calculating $\grf{\mathrm{d}\kappa/\mathrm{d}f}_\mathrm{K}^{\mathrm{\ton{pK}}}$;
instead, they vary instantaneously  because of the pK accelerations. Furthermore, the first term of \rfr{ugh}, calculated with ${\bds A}^\mathrm{\ton{1}}$ and ${\bds A}^\mathrm{\ton{2}}$ onto the Keplerian ellipse, respectively, yield the first two terms of \rfr{referee2}. The term in square brackets of \rfr{ugh}, calculated with ${\bds A}^\mathrm{\ton{2}}$ onto the Keplerian ellipse and multiplied by $\mathrm{d}\kappa/\mathrm{d}f$, calculated with ${\bds A}^\mathrm{\ton{1}}$ onto the Keplerian ellipse,  gives rise to the second mixed term of \rfr{eq2}, while the term in square brackets of \rfr{ugh}, calculated with ${\bds A}^\mathrm{\ton{1}}$ onto the Keplerian ellipse and multiplied by $\mathrm{d}\kappa/\mathrm{d}f$, calculated with ${\bds A}^\mathrm{\ton{2}}$ onto the Keplerian ellipse,  gives rise to the fourth mixed term of \rfr{eq2}.
}

It should be noted that  also effects \textcolor{black}{of order} $\mathcal{O}\ton{A^2}$ arise from \rfr{eq2} if it is calculated with the \textcolor{black}{same} pK acceleration.
%at a time, i.e. for $\ton{2}\rightarrow\ton{1}$ in the displayed expression, and $\ton{1}\rightarrow\ton{2}$ in the same with the exchange $\ton{1} %\leftrightarrows\,\ton{2}$ with respect to the previous one.
They will not be treated here since they would be \textcolor{black}{of order} $\mathcal{O}\ton{J_2^2}$ and  $\mathcal{O}\ton{1/c^4}$, respectively.
\subsection{The mixed averaged shifts per orbit \textcolor{black}{of order} $\mathcal{O}\ton{J_2/c^2}$}\lb{j2c2}
Let us assume that \citep{2014grav.book.....P}
\begin{align}
{\bds A}^\mathrm{\ton{1}} \lb{AccA} &\equiv {\bds A}^{\ton{J_2}} = \rp{3\,J_2\,\,R^2\,\mu}{2\,r^4}\qua{\ton{5\,\xi^2 - 1}\,\bds{\hat{r}} - 2\,\xi\,\bds{\hat{k}} }, \acap
{\bds A}^\mathrm{\ton{2}} \lb{AccB} &\equiv {\bds A}^{\ton{\mathrm{1pN}}} = \rp{\mu}{c^2\,r^2}\qua{\ton{\rp{4\,\mu}{r} - v^2}\,\bds{\hat{r}} + 4\,v_r\,\bds{v} },
\end{align}
where $R$ is the body's equatorial radius, $\bds{\hat{k}}$ is the unit vector directed along its symmetry axis,
\eqi
\xi:= \bds{\hat{k}}\bds\cdot\bds{\hat{r}}
\eqf
is the cosine of the angle between the body's spin axis and the satellite's position vector,
%$c$ is the speed of light in vacuum,
and
\eqi
v_r := \bds{v}\bds\cdot\bds{\hat{r}}
\eqf
is the radial velocity of the test particle.
It turns out that the $R-T-N$ components of the accelerations of \rfrs{AccA}{AccB} are
\begin{align}
A_R^{\ton{J_2}} \lb{ARJ2}& =\rp{3\,J_2\,R^2\mu\,\ton{1 + e\,\cos f}^4\,\qua{-1 + 3\,\ton{\kl\,\cos u + \km\,\sin u}^2}}{2\,a^4\,\ton{1 - e^2}^4}, \acap
A_T^{\ton{J_2}} \lb{ATJ2}& = -\rp{3\,J_2\,R^2\mu\,\ton{1 + e\,\cos f}^4\,\ton{\kl\,\cos u + \km\,\sin u}\,\ton{-\kl\,\sin u +
   \km\,\cos u}}{a^4\,\ton{1 - e^2}^4}, \acap
A_N^{\ton{J_2}} \lb{ANJ2}& = -\rp{3\,J_2\,R^2\mu\,\ton{1 + e\,\cos f}^4\,\kh\,\ton{\kl\,\cos u + \km\,\sin u}}{a^4\,\ton{1 - e^2}^4},
\end{align}
and
\begin{align}
A_R^{\ton{\mathrm{1pN}}} \lb{ARpN} & = -\rp{\mu^2\,\ton{1 + e\,\cos f}^2\,\ton{-3 - e^2 - 2\,e\,\cos f + 2\,e^2\,\cos 2 f}}{c^2\,a^3\,\ton{1 - e^2}^3}, \acap
A_T^{\ton{\mathrm{1pN}}} \lb{ATpN} & =\rp{4\,e\,\mu^2\,\ton{1 + e\,\cos f}^3\,\sin f}{c^2\,a^3\,\ton{1 - e^2}^3}, \acap
A_N^{\ton{\mathrm{1pN}}} \lb{ANpN} & =0.
\end{align}
In \rfrs{ARJ2}{ANJ2},
\eqi
\bds{\hat{l}} = \grf{\cos\Omega,~\sin\Omega,~0}
\eqf
is the unit vector directed along the line of the nodes toward the ascending node,
\eqi
\bds{\hat{m}}=\grf{-\cos I\sin\Omega,~\cos I\cos\Omega,~\sin I}\
\eqf
is the unit vector directed transversely to the line of the nodes in the orbital plane,
and
\eqi
\bds{\hat{h}}=\grf{\sin I\sin\Omega,~-\sin I\cos\Omega,~\,\cos I}
\eqf
is the normal unit vector, directed along the orbital angular momentum, such that $\bds{\hat{l}}\bds\times\bds{\hat{m}}=\bds{\hat{h}}$; see, e.g., \citet{1991ercm.book.....B,Sof89}.

By inserting \rfrs{ARJ2}{ANJ2} in \rfrs{dadt}{dodt}, one can use \rfr{Dkappa}  and \rfr{dtdf} to calculate the $J_2$-driven \textrm{instantaneous} orbital shifts needed in \rfr{eq2}. They turn out to be
\begin{align}
\Delta a^{\ton{J_2}}\ton{f}\lb{DaJ2f}& =  -\rp{J_2\,R^2}{16\,a\,\ton{1 - e^2}^3}\,
\textcolor{black}{
\mathcal{A}^{\ton{J_2}},
}
\acap
\Delta e\textcolor{black}{^{\ton{J_2}}}\ton{f} & = \rp{J_2\,R^2}{32\,a^2\,\ton{1 - e^2}^2}\,
\textcolor{black}{
\mathcal{E}^{\ton{J_2}},
}
\acap
\Delta I\textcolor{black}{^{\ton{J_2}}}\ton{f} & = -\rp{J_2\,R^2}{4\,a^2\,\ton{1 - e^2}^2}\,
\textcolor{black}{
\mathcal{I}^{\ton{J_2}},
}
\acap
\Delta \Omega\textcolor{black}{^{\ton{J_2}}}\ton{f} & = -\rp{J_2\,R^2\,\csc I}{4\,a^2\,\ton{1 - e^2}^2}\,
\textcolor{black}{
\mathcal{N}^{\ton{J_2}},
}
\acap
\Delta \omega\textcolor{black}{^{\ton{J_2}}}\ton{f} \lb{DoJ2f}& = \rp{J_2\,R^2}{32\,a^2\,e\,\ton{1 - e^2}^2}\,
\textcolor{black}{
\mathcal{P}^{\ton{J_2}},
}
\end{align}
\textcolor{black}{where
\begin{align}
\mathcal{A}^{\ton{J_2}} \lb{c_a_J2} & :=\sum_{j=1}^6\mathcal{A}^{\ton{J_2}}_j\,\widehat{T}_j, \acap
\mathcal{E}^{\ton{J_2}} & :=\sum_{j=1}^6\mathcal{E}^{\ton{J_2}}_j\,\widehat{T}_j, \acap
\mathcal{I}^{\ton{J_2}} & :=\sum_{j=1}^6\mathcal{I}^{\ton{J_2}}_j\,\widehat{T}_j, \acap
\mathcal{N}^{\ton{J_2}} & :=\sum_{j=1}^6\mathcal{N}^{\ton{J_2}}_j\,\widehat{T}_j, \acap
\mathcal{P}^{\ton{J_2}} \lb{c_o_J2} & :=\sum_{j=1}^6\mathcal{P}^{\ton{J_2}}_j\,\widehat{T}_j.
\end{align}
The coefficients $\widehat{T}_j,\,j=1,\,2,\,\dots 6$ are displayed in Appendix\,\ref{appendice}, while $\mathcal{A}_1^{\ton{J_2}},\ldots \mathcal{P}_6^{\ton{J_2}}$  are shown in Appendix\,\ref{appenA1}.
}
\Rfr{Dkappa}, calculated with \rfrs{ARpN}{ANpN} in \rfrs{dadt}{dodt} and \rfr{dtdf}, yields for the 1pN \textrm{instantaneous} orbital shifts
\begin{align}
\Delta a^{\mathrm{\ton{1pN}}}\ton{f}\lb{DapNf}& = -\rp{2\,e\,\mu\,\ton{\cos f- \cos f_0}\,\qua{7 + 3\,e^2 + 5\,e\,\ton{\cos f + \cos f_0}}}{c^2\,\ton{1 - e^2}^2},\acap
\Delta\,e^{\mathrm{\ton{1pN}}}\ton{f}\lb{DepNf}& = -\rp{\mu\,\ton{\cos f- \cos f_0}\,\qua{3 + 7\,e^2 + 5\,e\,\ton{\cos f + \cos f_0}}}{c^2\,a\,\ton{1 - e^2}},\acap
\Delta I^{\mathrm{\ton{1pN}}}\ton{f}\lb{DIpNf}& =0,\acap
\Delta \Omega^{\mathrm{\ton{1pN}}}\ton{f}\lb{DOpNf}& =0,\acap
\Delta \omega^{\mathrm{\ton{1pN}}}\ton{f}\lb{DopNf}& = -\rp{\mu\,\grf{
3\,e\,\ton{-f + f_0} + \ton{3 - e^2 + 5\,e\,\cos f}\,\sin f + \ton{-3 + e^2 - 5\,e\,\cos f_0}\,\sin f_0
}}{c^2\,a\,e\,\ton{1 - e^2}}.
\end{align}
The instantaneous shifts of $\eta$ are not displayed since they are not required in \rfr{eq2}. Indeed,  \textcolor{black}{as pointed out in Section\,\ref{over},} the derivatives $\mathrm{d}\kappa/\mathrm{d}f$, calculated with either \rfrs{ARJ2}{ANJ2} or \rfrs{ARpN}{ANpN} in \rfrs{dadt}{detadt}, do not contain explicitly
%\footnote{Note that it is a general feature; indeed, the summation over the Keplerian orbital elements in \rfr{eq2} does not include $\eta$.}
$\eta$.

\Rfr{eq2}, applied to \rfrs{AccA}{AccB} and calculated with \rfrs{DaJ2f}{DoJ2f} and \rfrs{DapNf}{DopNf}, allows to obtain the total mixed  \textrm{shifts per orbit} \textcolor{black}{of order} $\mathcal{O}\ton{J_2/c^2}$. They are
\begin{align}
\Delta a^{\ton{J_2/c^2}}_\mathrm{mix} \lb{Da}&= \rp{9\,\pi\,J_2\,R^2\,\mu}{4\,c^2\,a^2\,\ton{1-e^2}^4}\,\mathcal{\overline{A}}^{\textcolor{black}{\ton{J_2/c^2}}}\textcolor{black}{,}\acap
\Delta\,e^{\ton{J_2/c^2}}_\mathrm{mix} \lb{De}&= -\rp{3\,\pi\,J_2\,R^2\,\mu}{8\,c^2\,a^3\,\ton{1-e^2}^3}\,\mathcal{\overline{E}}^{\textcolor{black}{\ton{J_2/c^2}}}\textcolor{black}{,}\acap
\Delta I^{\ton{J_2/c^2}}_\mathrm{mix} \lb{DI}&= -\rp{3\,\pi\,J_2\,R^2\,\mu}{c^2\,a^3\,\ton{1-e^2}^3}\,\mathcal{\overline{I}}^{\textcolor{black}{\ton{J_2/c^2}}}\textcolor{black}{,}\acap
\Delta \Omega^{\ton{J_2/c^2}}_\mathrm{mix} \lb{DO}&= -\rp{3\,\pi\,J_2\,R^2\,\mu\,\csc I}{c^2\,a^3\,\ton{1-e^2}^3}\,\mathcal{\overline{N}}^{\textcolor{black}{\ton{J_2/c^2}}}\textcolor{black}{,}\acap
\Delta \omega^{\ton{J_2/c^2}}_\mathrm{mix} \lb{Do}&= -\rp{3\,\pi\,J_2\,R^2\,\mu\,\csc I}{16\,c^2\,a^3\,e\,\ton{1-e^2}^3}\,\mathcal{\overline{P}}^{\textcolor{black}{\ton{J_2/c^2}}}\textcolor{black}{,}\acap
\Delta \eta^{\ton{J_2/c^2}}_\mathrm{mix} \lb{Deta}&= -\rp{3\,\pi\,J_2\,R^2\,\mu}{16\,c^2\,a^3\,e\,\ton{1-e^2}^{7/2}}\,\mathcal{\overline{H}}^{\textcolor{black}{\ton{J_2/c^2}}},
\end{align}
where
\textcolor{black}{
\begin{align}
\mathcal{\overline{A}}^{\ton{J_2/c^2}} \lb{coeff_sma} & :=\sum_{j = 1}^6\mathcal{\overline{A}}^{\ton{J_2/c^2}}_j\,\widehat{T}_j, \\ \nonumber \\
\mathcal{\overline{E}}^{\ton{J_2/c^2}} \lb{coeff_ecce} & :=\sum_{j = 1}^6\mathcal{\overline{E}}^{\ton{J_2/c^2}}_j\,\widehat{T}_j, \\ \nonumber \\
\mathcal{\overline{I}}^{\ton{J_2/c^2}} \lb{coeff_incli} & :=\sum_{j = 1}^6\mathcal{\overline{I}}^{\ton{J_2/c^2}}_j\,\widehat{T}_j, \\ \nonumber \\
\mathcal{\overline{N}}^{\ton{J_2/c^2}} \lb{coeff_nodo} & :=\sum_{j = 1}^6\mathcal{\overline{N}}^{\ton{J_2/c^2}}_j\,\widehat{T}_j, \\ \nonumber \\
\mathcal{\overline{P}}^{\ton{J_2/c^2}} \lb{coeff_peri} & :=\sum_{j = 1}^6\mathcal{\overline{P}}^{\ton{J_2/c^2}}_j\,\widehat{T}_j, \\ \nonumber \\
\mathcal{\overline{H}}^{\ton{J_2/c^2}} \lb{coeff_eta} & :=\sum_{j = 1}^6\mathcal{\overline{H}}^{\ton{J_2/c^2}}_j\,\widehat{T}_j.
\end{align}
The explicit forms of the coefficients $\mathcal{\overline{A}}_1^{\ton{J_2/c^2}},\ldots \mathcal{\overline{H}}_6^{\ton{J_2/c^2}}$ entering \rfrs{coeff_sma}{coeff_eta} are displayed in Appendix\,\ref{appenA2}.
}
\section{The mixed averaged precessions}\lb{rates}
In calculating the mixed averaged orbital \textrm{precessions} $\dot\kappa^{\ton{J_2/c^2}}_\mathrm{mix}$, caution is needed. Their full expressions do not consist only of the ratios
\eqi
\rp{\Delta\kappa_\mathrm{mix}^{\ton{J_2/c^{-2}}}}{\Pb^{\ton{\mathrm{K}}}}
\eqf
of the mixed net shifts per orbit of \rfrs{Da}{Deta} to the Keplerian orbital period. Indeed, one has to include also the ratios of the direct averaged variations $\Delta\kappa^\mathrm{\ton{pK}}$ of the orbital elements due to a given pK acceleration ${\bds A}^\mathrm{\ton{pK}}$ to the total pK period
\eqi
\Pb^{\ton{\mathrm{pK}}} = \Pb^{\ton{\mathrm{K}}} + \Delta\Pb^{\ton{\mathrm{pK}}}
\eqf
including the corrections $\Delta\Pb^{\ton{\mathrm{pK}}}$ to $\Pb^{\ton{\mathrm{K}}}$ due to ${\bds A}^{\ton{1}}$ and  ${\bds A}^{\ton{2}}$, and expanding the resulting expressions to the order required to have just the desired mixed effects. Also in this case, the contributions \textcolor{black}{of order} $\mathcal{O}\ton{A^2}$ are neglected.

By following the calculational approach of
%\footnote{While \citet{2016MNRAS.460.2445I} obtained just \rfr{PpN} for $\Delta\Pb^{\mathrm{\ton{1pN}}}$, only an approximated expression to the zero %order in $e$ was displayed in \citet{2016MNRAS.460.2445I} for $\Delta\Pb^{\ton{J_2}}$; it agrees with the limit of \rfr{PJ2} for $e\rightarrow 0$.}
\citet{2016MNRAS.460.2445I} for the anomalistic orbital period, it turns out that the corrections to $\Pb^{\ton{\mathrm{K}}}$ induced by \rfrs{AccA}{AccB} are
\begin{align}
\Delta\Pb^{\ton{J_2}}  \lb{PJ2}& = \rp{3\,\pi\,J_2\,R^2\,\ton{1 + e\,\cos f_0}^3}{2\,\sqrt{\mu\,a}\,\ton{1 - e^2}^3}\,\textcolor{black}{\qua{-2 + 3\,\ton{\widehat{T}_2 + \widehat{T}_3\,\cos 2u_0} + 6\,\widehat{T}_6\,\sin 2u_0}}, \\ \nonumber\\
\Delta\Pb^{\ton{\mathrm{1pN}}} \lb{PpN}& = \rp{3\,\pi\,\sqrt{\mu\,a}}{c^2\,\ton{1 - e^2}^2}\,\qua{6 + 7\,e^2 + 2\,e^4 + 2\,e\,\ton{7 + 3\,e^2}\,\cos f_0 +
5\,e^2\,\cos 2f_0},
\end{align}
respectively.

The \textrm{averaged} orbital changes due to $J_2$
are
\begin{align}
\Delta a^{\ton{J_2}} \lb{DaJ2}& = 0, \acap
\Delta\,e^{\ton{J_2}} \lb{DeJ2}& = 0, \acap
\Delta I^{\ton{J_2}} \lb{DIJ2}& = -\rp{3\,\pi\,J_2\,R^2\,\textcolor{black}{\widehat{T}_4}}{a^2\,\ton{1 - e^2}^2}, \acap
\Delta \Omega^{\ton{J_2}} \lb{DOJ2}& = -\rp{3\,\pi\,J_2\,R^2\,\csc I\,\textcolor{black}{\widehat{T}_5}}{a^2\,\ton{1 - e^2}^2}, \acap
\Delta \omega^{\ton{J_2}} \lb{DoJ2}& = \rp{3\,\pi\,J_2\,R^2\,\ton{2 - 3\,\textcolor{black}{\widehat{T}_2} + 2\,\textcolor{black}{\widehat{T}_5}\,\cot I}}{2\,a^2\,\ton{1 - e^2}^2}, \acap
\Delta \eta^{\ton{J_2}} \lb{DetaJ2}& = \rp{3\,\pi\,J_2\,R^2\,\ton{2 - 3\,\textcolor{black}{\widehat{T}_2}}}{2\,a^2\,\ton{1 - e^2}^{3/2}}.
\end{align}

The 1pN \textrm{averaged} orbital shifts are
\begin{align}
\Delta a^{\ton{\mathrm{1pN}}} \lb{DapN}& = 0, \acap
\Delta\,e^{\ton{\mathrm{1pN}}} \lb{DepN}& = 0, \acap
\Delta I^{\ton{\mathrm{1pN}}} \lb{DIpN}& =0, \acap
\Delta \Omega^{\ton{\mathrm{1pN}}} \lb{DOpN}& = 0, \acap
\Delta \omega^{\ton{\mathrm{1pN}}} \lb{DopN}& = \rp{6\,\pi\,\mu}{c^2\,a\,\ton{1-e^2}}, \acap
\Delta \eta^{\ton{\mathrm{1pN}}} \lb{DetapN}& = \rp{6\,\pi\,\mu}{c^2\,a}\,\ton{2 -\rp{5}{\sqrt{1-e^2}}}.
\end{align}
\Rfrs{DaJ2}{DoJ2} and \rfrs{DapN}{DopN} are calculated by posing $f = f_0 + 2\pi$ in \rfrs{DaJ2f}{DoJ2f} and \rfrs{DapNf}{DopNf}, respectively.
\Rfr{DetaJ2} and \rfr{DetapN} can straightforwardly be obtained from Eq.\,(48) and Eq.\,(29), respectively, of \citet{2019EPJC...79..816I} by multiplying the latter ones by $\Pb^{\mathrm{\ton{K}}} = 2\pi/\nk^{\mathrm{\ton{K}}}$.

By using \rfr{PpN} with \rfrs{DaJ2}{DetaJ2} and \rfr{PJ2} with \rfrs{DapN}{DetapN}, one finally gets
\begin{align}
\dot a_\mathrm{mix}^{\ton{J_2/c^{-2}}} \lb{dota_mix}& = 0,\acap
\dot e_\mathrm{mix}^{\ton{J_2/c^{-2}}} \lb{dote_mix}& = 0,\acap
\dot I_\mathrm{mix}^{\ton{J_2/c^{-2}}} \lb{dotI_mix}& = \rp{9\,\nk^{\ton{\mathrm{K}}}\,J_2\,R^2\,\mu\,\textcolor{black}{\widehat{T}_4}}{4\,c^2\,a^3\,\ton{1 - e^2}^4}\,\qua{6 + 7\,e^2 + 2\,e^4 + 2\,e\,\ton{7 + 3\,e^2}\,\cos f_0 + 5\,e^2\,\cos 2f_0},\acap
\dot \Omega_\mathrm{mix}^{\ton{J_2/c^{-2}}} \lb{dotO_mix}& = \rp{9\,\nk^{\ton{\mathrm{K}}}\,J_2\,R^2\,\mu\,\textcolor{black}{\widehat{T}_5}\,\csc I}{4\,c^2\,a^3\,\ton{1 - e^2}^4}\,\qua{6 + 7\,e^2 + 2\,e^4 + 2\,e\,\ton{7 + 3\,e^2}\,\cos f_0 + 5\,e^2\,\cos 2f_0},\acap
\dot \omega_\mathrm{mix}^{\ton{J_2/c^{-2}}} \lb{doto_mix}\nonumber & = \rp{9\,\nk^{\ton{\mathrm{K}}}\,J_2\,R^2\,\mu}{8\,c^2\,a^3\,\ton{1 - e^2}^4}\,\grf{
\qua{6 + 7\,e^2 + 2\,e^4 + 2\,e\,\ton{7 + 3\,e^2}\,\cos f_0 + 5\,e^2\,\cos 2f_0}\times\right.\acap
&\left.\times\,\ton{-2 + 3\,\textcolor{black}{\widehat{T}_2} - 2\,\textcolor{black}{\widehat{T}_5}\,\cot I} + 2\,\ton{1 + e\,\cos f_0}^3\,\qua{2 - 3\,\ton{\textcolor{black}{\widehat{T}_2} \textcolor{black}{+} \textcolor{black}{\widehat{T}_3}\,\cos 2 u_0} - 6\,\textcolor{black}{\widehat{T}_6}\,\sin 2u_0}
},\acap
\dot \eta_\mathrm{mix}^{\ton{J_2/c^{-2}}} \lb{doteta_mix}\nonumber & = -\rp{9\,\nk^{\ton{\mathrm{K}}}\,J_2\,R^2\,\mu}{8\,c^2\,a^3\,\ton{1 - e^2}^{7/2}}\,\grf{
\ton{2 - 3\,\textcolor{black}{\widehat{T}_2}}\,\qua{6 + 7\,e^2 + 2\,e^4 + 2\,e\,\ton{7 + 3\,e^2}\,\cos f_0 + 5\,e^2\cos 2f_0} +\right.\acap
&\left. + 2\,\ton{5 - 2\,\sqrt{1 - e^2}}\,\ton{1 + e\,\cos f_0}^3\,\qua{2 - 3\,\ton{\textcolor{black}{\widehat{T}_2 + \widehat{T}_3}\,\cos 2u_0} - 6\,\textcolor{black}{\widehat{T}_6}\,\sin 2u_0}
}.
\end{align}
\textcolor{black}{In \rfrs{doto_mix}{doteta_mix}, it is $u_0 :=f_0 + \omega $.}
\Rfrs{dota_mix}{doteta_mix} add to the ratios of \rfrs{Da}{Deta} to the Keplerian orbital period $\Pb^{\mathrm{\ton{K}}}$ in order to give the \textrm{total} mixed orbital \textrm{precessions} \textcolor{black}{of order} $\mathcal{O}\ton{J_2/c^2}$; the resulting expressions are too cumbersome to be shown here.

\textcolor{black}{
\Rfrs{dota_mix}{doteta_mix} represent, together
with the results of Section\,\ref{shifts}, the primary findings of this investigation. Some special
configurations will be investigated in some more detail in the next Section\,\ref{orbi}.
}
\section{Some special orbital configurations}\lb{orbi}
Two peculiar orbital configurations are considered here: a) equatorial (Section\,\ref{equa}) and b) polar (Section\,\ref{pola}) orbits.

By parameterizing $\bds{\hat{k}}$ in terms of the right ascension (RA) $\alpha$ and declination (DEC) $\delta$ of the body's north pole of rotation as
\eqi
\bds{\hat{k}} =\grf{\cos\alpha\,\cos\delta,\,\sin\alpha\,\cos\delta,\,\sin\delta},\lb{kvers}
\eqf
%as
%\begin{align}
%{\hat{k}}_x \lb{kx} & = \cos\alpha\cos\delta, \\ \nonumber \\
%
%{\hat{k}}_y \lb{ky} & = \sin\alpha\cos\delta, \\ \nonumber \\
%
%{\hat{k}}_z \lb{kz} & = \sin\delta, \\ \nonumber \\
%
%\end{align}
one obtains
\begin{align}
\kl \lb{kl}& = \cos\delta\,\cos\ton{\alpha - \Omega}, \\ \nonumber \\
\km \lb{km}& = \sin I\,\sin\delta + \cos I \,\cos\delta\,\sin\ton{\alpha - \Omega}, \\ \nonumber \\
\kh \lb{kh}& = \cos I\,\sin\delta - \sin I\,\cos\delta\,\sin\ton{\alpha - \Omega}.
\end{align}
\textcolor{black}{
These expressions will be used in what follows.
}
\subsection{Equatorial orbit}\lb{equa}
Let us assume that the satellite's orbital plane lies in the equatorial plane of the primary,  whatever the orientation of the latter in the adopted reference frame, i.e., for generic values of $\alpha,\,\delta$: for such an orbital geometry, it is
\begin{align}
\kl \lb{c1}& = \km =0, \\ \nonumber \\
\kh \lb{c2}& =1.
\end{align}
According to \rfrs{kl}{kh}, the conditions of \rfrs{c1}{c2} are satisfied if
\begin{align}
I \lb{I1}& = \rp{\pi}{2} - \delta, \\ \nonumber \\
\Omega \lb{O1}& = \alpha + \rp{\pi}{2}.
\end{align}

Then, \rfrs{Da}{Deta} reduce to
\begin{align}
\Delta a_\mathrm{mix}^{\ton{J_2/c^2}} & = 0,\acap
\Delta e_\mathrm{mix}^{\ton{J_2/c^2}} & = 0,\acap
\Delta I_\mathrm{mix}^{\ton{J_2/c^2}} & = 0,\acap
\Delta \Omega_\mathrm{mix}^{\ton{J_2/c^2}} & = 0,\acap
\Delta \omega_\mathrm{mix}^{\ton{J_2/c^2}} & = \rp{3\,\pi\,J_2\,R^2\,\mu\,\ton{44 + 17\,e^2 - 64\,e\,\cos f_0}}{4\,c^2\,a^3\,\ton{1 - e^2}^3},\acap
\Delta \eta_\mathrm{mix}^{\ton{J_2/c^2}} \nonumber & = \rp{3\,\pi\,J_2\,R^2\,\mu\,}{4\,c^2\,a^3\,\ton{1 - e^2}^{7/2}}\,\ton{
-88 + 16\,\sqrt{1 - e^2} + e^2\,\qua{63 - 5\,e^2 + 24\,\sqrt{1 - e^2}} +\right.\acap
\beg\left. + e\,\grf{3\,e^2\,\qua{7 + 4\,\sqrt{1 - e^2}} + 8\,\qua{-17 + 6\,\sqrt{1 - e^2}}}\,\cos f_0 + \right.\acap
&\left. + 8\,e^2\,\qua{-5 + 3\,\sqrt{1 - e^2}}\,\cos 2 f_0 + e^3\,\qua{-5 + 4\,\sqrt{1 - e^2}}\,\cos 3 f_0
},
\end{align}
while \rfrs{dota_mix}{doteta_mix} become
\begin{align}
\dot a_\mathrm{mix}^{\ton{J_2/c^2}} & = 0,\acap
\dot e_\mathrm{mix}^{\ton{J_2/c^2}} & = 0,\acap
\dot I_\mathrm{mix}^{\ton{J_2/c^2}} & = 0,\acap
\dot \Omega_\mathrm{mix}^{\ton{J_2/c^2}} & = 0,\acap
\dot \omega_\mathrm{mix}^{\ton{J_2/c^2}} & = -\rp{9\,\nk^{\mathrm{\ton{K}}}\,J_2\,R^2\,\mu\,\qua{
8 + 8\,e^2 + 4\,e^4 + e\,(16 + 9\,e^2)\,\cos f_0 + 4\,e^2\,\cos 2 f_0 - e^3\,\cos 3 f_0
}}{8\,c^2\,a^3\,\ton{1 - e^2}^4},
\acap
\dot \eta_\mathrm{mix}^{\ton{J_2/c^2}} \nonumber & = \rp{9\,\nk^{\mathrm{\ton{K}}}\,J_2\,R^2\,\mu}{4\,c^2\,a^3\,\ton{1 - e^2}^{7/2}}\,\qua{
6 + 7\,e^2 + 2\,e^4 + 2\,e\,\ton{7 + 3\,e^2}\,\cos f_0 + \right.\acap
&\left. + 2\,\ton{5 - 2\,\sqrt{1 - e^2}}\,\ton{1 + e\,\cos f_0}^3 + 5\,e^2\,\cos 2 f_0
}.
\end{align}
\subsection{Polar orbit}\lb{pola}
Let us, now, assume that the body's spin axis, irrespectively of its orientation, i.e., for generic values of $\alpha,\,\delta$, lies somewhere in the satellite's orbital plane between $\bds{\hat{l}}$ and $\bds{\hat{m}}$.  In such a scenario, it is
\begin{align}
\kl \lb{pel}& \neq 0, \\ \nonumber \\
\km \lb{pm}& \neq 0, \\ \nonumber \\
\kh \lb{ph}& =0.
\end{align}
According to \rfrs{kl}{kh}, the conditions of \rfrs{pel}{ph} are fulfilled if
\begin{align}
I \lb{inc}& = \rp{\pi}{2}, \\ \nonumber \\
\Omega \lb{omg}& = \alpha;
\end{align}
indeed, with \rfrs{inc}{omg}, one has just
\begin{align}
\kl & =\cos\delta, \\ \nonumber \\
\km & =\sin\delta, \\ \nonumber \\
\kh & =0.
\end{align}

Thus, \rfrs{Da}{Deta} reduce to
\begin{align}
\Delta a_\mathrm{mix}^{\ton{J_2/c^2}} \nonumber & = -\rp{9\,\pi\,J_2\,R^2\,\mu}{4\,c^2\,a^2\,\ton{1 - e^2}^4}\,\grf{
e^3\,\sin\ton{f_0 + 2 \delta - 2 \omega} + e^2\,\ton{12 + e^2}\,\sin\ton{2\delta - 2\omega} -\right.\acap
\beg\left. -  2\,\qua{4 + 6\,e^2\,+ 3\,e\,\ton{4 + e^2}\,\cos f_0}\,\sin\ton{2f_0 - 2\delta + 2\omega} -
 6\,e^2\,\sin\ton{4f_0 - 2\delta + 2\omega} - \right.\acap
 &\left. - e^3\,\sin\ton{5 f_0 - 2 \delta + 2 \omega}
}, \acap
\Delta e_\mathrm{mix}^{\ton{J_2/c^2}} \nonumber & = \rp{3\,\pi\,J_2\,R^2\,\mu}{8\,c^2\,a^3\,\ton{1 - e^2}^3}\,\grf{
4\,\qua{3\,\sin\ton{f_0 - 2 \delta + 2 \omega} + 7\,\sin\ton{3 f_0 - 2 \delta + 2 \omega}} +\right.\acap
\beg\left. + e\,\qua{-3\,e\,\sin\ton{f_0 + 2 \delta - 2 \omega} - \ton{20 + 19\,e^2}\,\sin\ton{2\delta - 2\omega} + 60\,\sin\ton{2f_0 - 2\delta + 2\omega} +\right.\right.\acap
\beg\left.\left. + 18\,\sin\ton{4f_0 - 2\delta + 2\omega} + 33\,e\,\sin\ton{f_0 - 2 \delta + 2 \omega} + 17\,e\,\sin\ton{3 f_0 - 2 \delta + 2 \omega} +\right.\right.\acap
&\left.\left. + 3\,e\,\sin\ton{5 f_0 - 2 \delta + 2 \omega}
}
}, \acap
\Delta I_\mathrm{mix}^{\ton{J_2/c^2}}  & = 0, \acap
\Delta \Omega_\mathrm{mix}^{\ton{J_2/c^2}}  & = 0, \acap
\Delta \omega_\mathrm{mix}^{\ton{J_2/c^2}} \nonumber & = -\rp{3\,\pi\,J_2\,R^2\,\mu}{8\,c^2\,a^3\,e\,\ton{1 - e^2}^3}\,\ton{
\ton{- 12 + 45\,e^2}\,\cos\ton{f_0 - 2 \delta + 2 \omega} + \ton{28 + 19\,e^2}\,\cos\ton{3 f_0 - 2 \delta + 2 \omega} + \right.\acap
\beg\left. + e\,\grf{2\,\ton{-10 + 9\,e^2}\,\cos\ton{2\delta - 2\omega} + 60\,\cos\ton{2f_0 - 2\delta + 2\omega} + \right.\right.\acap
\nonumber &\left.\left. + 18\,\cos\ton{4f_0 - 2\delta + 2\omega} + 3\,e\,\cos\ton{5 f_0 - 2 \delta + 2 \omega}+\right.\right.\acap
&\left.\left.  + 44 + 17\,e^2 - e\,\qua{64\,\cos f_0 + 3\,\cos\ton{f_0 + 2\delta - 2 \omega}}}
},
\acap
\Delta \eta_\mathrm{mix}^{\ton{J_2/c^2}} \nonumber & = -\rp{3\,\pi\,J_2\,R^2\,\mu}{16\,c^2\,a^3\,e\,\ton{1 - e^2}^{7/2}}\,\qua{
-2\,e\,\grf{88 + 5\,e^4 - 16\,\sqrt{1 - e^2} - 3\,e^2\,\ton{21 + 8\,\sqrt{1 - e^2}} - \right.\right.\acap
\nonumber &\left.\left. - e\,\qua{3\,e^2\,\ton{7 + 4\,\sqrt{1 - e^2}} + 8\,\ton{-17 + 6\,\sqrt{1 - e^2}}}\,\cos f_0 + \right.\right.\acap
\nonumber &\left.\left. + e^2\,\qua{8\,\ton{5 - 3\,\sqrt{1 - e^2}}\,\cos 2f_0 + e\,\ton{5 - 4\,\sqrt{1 - e^2}}\,\cos 3f_0}} +\cos 2\delta\,\grf{
3\,e^2\,\ton{2 - 7\,e^2}\,\cos\ton{f_0 - 2\omega} - \right.\right.\acap
\nonumber &\left.\left. - 2\,e\,\ton{-20 + 7\,e^2 + 13\,e^4}\,\cos 2\omega + 12\,e\,\qua{-14 - 11\,e^2 + 8\,\sqrt{1 - e^2}\,\ton{1 + e\,\cos f_0}^3}\,\cos 2u_0 - \right.\right.\acap
\nonumber &\left.\left. - 18\,e\,\ton{2 + 3\,e^2}\,\cos\ton{4f_0 + 2\omega} - 3\,\ton{-8 + 74\,e^2 + 9\,e^4}\,\cos\ton{f_0 + 2\omega} - \right.\right.\acap
\nonumber &\left.\left. - \ton{4 + e^2}\,\ton{14 + 31\,e^2}\,\cos\ton{3 f_0 + 2\omega} - 3\,e^2\,\ton{2 + 3\,e^2}\,\cos\ton{5f_0 + 2\omega}
}
-\right.\acap
\nonumber &\left. - \sin 2\delta\,\ton{3\,e^2\,\qua{2 + e^2\,\ton{-7 + 4\,\sqrt{1 - e^2}}}\,\sin\ton{f_0 - 2\omega} +
2 \,e\,\qua{-20 + 13\,e^4 + e^2\,\ton{7 - 36\,\sqrt{1 - e^2}}}\,\sin 2\omega + \right.\right.\acap
\nonumber &\left.\left. + 12\,e\,\qua{14 - 8\,\sqrt{1 - e^2} + e^2\,\ton{11 - 12\sqrt{1 - e^2}}}\,\sin 2u_0 +
18\,e\,\qua{2 + e^2\,\ton{3 - 4\,\sqrt{1 - e^2}}}\,\sin\ton{4f_0 + 2\omega} + \right.\right.\acap
\nonumber &\left.\left. + \grf{-24 + 3\,e^2\,\qua{74 - 48\,\sqrt{1 - e^2} + e^2\,\ton{9 - 12\sqrt{1 - e^2}}}}\,\sin\ton{f_0 + 2\omega} + \right.\right.\acap
\nonumber &\left.\left. + \grf{56 + e^2\,\qua{138 - 144\,\sqrt{1 - e^2} + e^2\,\ton{31 - 36\sqrt{1 - e^2}}}}\,\sin\ton{3f_0 + 2\omega} + \right.\right.\acap
&\left.\left. + 3\,e^2\,\qua{2 + e^2\,\ton{3 - 4\sqrt{1 - e^2}}}\,\sin\ton{5f_0 + 2\omega}
}
},
\end{align}
while \rfrs{dota_mix}{doteta_mix} can be written as
\begin{align}
\dot a_\mathrm{mix}^{\ton{J_2/c^2}} & = 0,\acap
\dot e_\mathrm{mix}^{\ton{J_2/c^2}} & = 0,\acap
\dot I_\mathrm{mix}^{\ton{J_2/c^2}} & = 0,\acap
\dot \Omega_\mathrm{mix}^{\ton{J_2/c^2}} & = 0,\acap
\dot \omega_\mathrm{mix}^{\ton{J_2/c^2}} \nonumber & = \rp{9\,\nk^{\mathrm{\ton{K}}}\,J_2\,R^2\,\mu}{8\,c^2\,a^3\,\ton{1 - e^2}^4}\,\grf{
6 + 7\,e^2 + 2\,e^4 + 2\,e\,\ton{7 + 3\,e^2}\,\cos f_0 + 5\,e^2\,\cos 2f_0 - \right.\acap
&\left. - 2\,\ton{1 + e\,\cos f_0}^3\,\qua{1 + 3\,\cos\ton{2f_0 -2\delta + 2\omega}}
},\acap
\dot \eta_\mathrm{mix}^{\ton{J_2/c^2}} \nonumber & = -\rp{9\,\nk^{\mathrm{\ton{K}}}\,J_2\,R^2\,\mu}{8\,c^2\,a^3\,\ton{1 - e^2}^{7/2}}\,\grf{
-6 - 7\,e^2 - 2\,e^4 - 2\,e\,\ton{7 + 3\,e^2}\,\cos f_0 - 5\,e^2\,\cos 2f_0 + \right.\acap
&\left. + 2\,\ton{-5 + 2\,\sqrt{1 - e^2}}\,\ton{1 + e\,\cos f_0}^3\,\qua{1 + 3\,\cos\ton{2f_0 -2\delta + 2\omega}}
}.
\end{align}
\subsection{\textcolor{black}{Numerical estimates}}\lb{num}
\textcolor{black}{
Here, some order-of-magnitude evaluations of the size of the effects derived in the previous Sections\,\ref{equa} to \ref{pola} for some natural and artificial bodies in our solar system are given.
}

\textcolor{black}{
For the sake of definiteness, just the amplitude
\eqi
\dert\psi t:=\rp{\nk^\mathrm{\ton{K}}\,J_2\,R^2\,\mu}{c^2\,a^3}\lb{dpsidt}
\eqf
is calculated.
}

\textcolor{black}{
For Mercury and the Sun, \rfr{dpsidt} yields $5\times 10^{-4}$ microarcseconds per century $\ton{\mu\mathrm{as\,cty}^{-1}}$, while for an Earth's artificial satellite like, e.g., LAGEOS \citep{2019JGeod..93.2181P}, one has $\dot\psi = 0.3$ milliarcseconds per year $\ton{\mathrm{mas\,yr}^{-1}}$. For Juno \citep{2017SSRv..213....5B}, currently orbiting Jupiter, it is $\dot\psi = 0.04$ microarcseconds per year $\ton{\mu\mathrm{as\,yr}^{-1}}$. While these figures are only rough indications, they should make it clear how small these effects are. Suffice it to say that, according to the recent planetary ephemerides EPM2017 \citep{2018AstL...44..554P}, the current (formal) accuracy in constraining any possible anomalous perihelion precession of Mercury may be calculated to be $\simeq 10\,\mu\mathrm{as\,cty}^{-1}$ \citep{2019AJ....157..220I}. Furthermore, it is still debated if the satellites of the LAGEOS family were actually able to measure the Lense-Thirring signal of a few tens of mas yr$^{-1}$ \citep{2013NuPhS.243..180C} to the per cent level \citep{2013CEJPh..11..531R}.
}

\textcolor{black}{
It is unclear if such mixed effects, which do contribute to the overall orbital evolution, could be actually measurable independently of other dynamical features of motion. Indeed, they do not come from some new pK acceleration, still unmodelled in the softwares used worldwide to process astronomical and geodetic observations of interest. If so, it could be possible, at least in principle, to include it in the dynamical models and estimate some dedicated solve-for parameters in the usual least-square approach of real data reductions. On the other hand, the standard pK accelerations of \rfr{AccA} and \rfr{AccB} giving rise to the features of motion which are the subject of this paper are accurately modelled; thus, just very tiny signatures, due to the current level of mismodeling in \rfrs{AccA}{AccB}, would impact the post-fit residuals produced in data analyses. Given the already tiny magnitude of the nominal values of such effects, it is even more unlikely than their resulting mismodelled signals may leave any detectable trace.
}

\section{Summary and conclusions}\lb{fine}
To the 1pN order, the net orbital effects per orbit experienced by a test particle moving around an oblate body include not only those \textrm{directly} induced by the 1pN acceleration \textcolor{black}{of order} $\mathcal{O}\ton{J_2/c^2}$, but further ones as well, also of the same order, due to the simultaneous action of two standard pK accelerations: the Newtonian one caused by the quadrupole mass moment $J_2$ of the primary, and the 1pN gravitoelectric one causing the formerly anomalous perihelion precession of Mercury in the field of the Sun. Such \textrm{indirect} features of motion \textcolor{black}{of order} $\mathcal{O}\ton{J_2/c^2}$ arise because, during an orbital revolution, the orbital elements do not remain constant, being instantaneously displaced by each of the pK accelerations. Moreover, the orbital period over which the average is performed is, actually, the time interval between two successive passages at the pericenter which instantaneously moves because of the pK perturbations.

We presented a general approach to analytically calculate the mixed effects arising from the interplay of two pK accelerations, irrespectively of their physical origin. As a result, we, first, explicitly calculated the mixed \textrm{net shifts per orbit} of all the osculating Keplerian orbital elements to the order of $\mathcal{O}\ton{J_2/c^2}$. It turned out that all of them undergo generally non-vanishing changes of this type.

Then, we worked out their mixed averaged \textrm{rates}  elucidating that their \textrm{total} expressions can only  be obtained  if also the ratios of the \textrm{direct} net shifts per orbit due to each pK acceleration  to the \textrm{pK} orbital period are taken in addition to the ratios of the \textrm{mixed} ones to the \textrm{Keplerian} one. Also in this case, analytic expressions of general validity were derived: no approximations pertaining both the satellite's orbital geometry and the spatial orientation of the body's spin axis were adopted. It turned out that the semimajor axis and the eccentricity stay constant in the aforementioned calculation.

Subsequently, we obtained simplified expressions for all the mixed effects under consideration that are specialized to the equatorial and polar orbit scenarios. In the former case, only the pericenter and the mean anomaly at epoch undergo non-vanishing mixed variations. In the latter, while the mixed net shifts per orbit of the inclination and the node are zero, the precessional contributions due to the pK period are non-vanishing only for the pericenter and the mean anomaly at epoch.

\textcolor{black}{
The nominal size of the effects studied in this work is very tiny for various astronomical scenarios of potential interest in our solar system; as an example, the perihelion of Mercury would be impacted at the level of $\simeq 10^{-4}\,\mu\mathrm{as\,cty}^{-1}$, compared to today's (formal) accuracy in constraining any possible anomalous precession of just  $\simeq 10\,\mu\mathrm{as\,cty}^{-1}$. Furthermore, since they arise from the interplay of standard accelerations which are routinely modelled to a high level of accuracy in the softwares used in real data reductions, only mismodelled signatures, much smaller than the already tiny nominal ones, should affect the post-fit residuals produced in data analyses.
}
\begin{appendices}
\section{\textcolor{black}{Coefficients of some orbital variations}}\lb{appendice}
\renewcommand{\theequation}{A\arabic{equation}}
\setcounter{equation}{0}
\textcolor{black}{
The coefficients $\widehat{T}_j,\,j=1,\,2,\,\dots 6$ entering \rfrs{c_a_J2}{c_o_J2} and \rfrs{coeff_sma}{coeff_eta} in Section\,\ref{j2c2} are
}
\textcolor{black}{
\begin{align}
\widehat{T}_1 & := 1,\acap
\widehat{T}_2 & := \qua{\ton{\kl}^2 + \ton{\km}^2},\acap
\widehat{T}_3 & := \qua{\ton{\kl}^2 - \ton{\km}^2},\acap
\widehat{T}_4 & := \qua{\ton{\kh}\,\ton{\kl}},\acap
\widehat{T}_5 & := \qua{\ton{\kh}\,\ton{\km}},\acap
\widehat{T}_6 & := \qua{\ton{\kl}\,\ton{\km}}.
\end{align}
They depend on $I$ and $\Omega$, and on the polar angles in terms of which $\bds{\hat{k}}$ is parameterized; see, e.g., \rfrs{kvers}{kh}.
}
\subsection{\textcolor{black}{Coefficients of the instantaneous Newtonian shifts due to $J_2$}}\lb{appenA1}
\renewcommand{\theequation}{A.1.\arabic{equation}}
\setcounter{equation}{0}
\textcolor{black}{
Here, we deal with the instantaneous Newtonian shifts induced by $J_2$ calculated in Section\,\ref{j2c2}. We display the explicit expressions of the coefficients $\mathcal{A}_1^{\ton{J_2}},\ldots \mathcal{P}_6^{\ton{J_2}}$ entering \rfrs{c_a_J2}{c_o_J2}, which are
}
\textcolor{black}{
\begin{align}
\mathcal{A}_1^{\ton{J_2}} & := 4\,e\,\qua{-3\,\ton{4 + e^2}\,\cos f + e\,\ton{-6\,\cos 2f - e\,\cos 3f}}-\qua{f \rightarrow f_0},\acap
\mathcal{A}_2^{\ton{J_2}} & := 6\,e\,\qua{3\,\ton{4 + e^2}\,\cos f + e\,\ton{6\,\cos 2f + e\,\cos 3f}} - \qua{f \rightarrow f_0},\acap
\mathcal{A}_3^{\ton{J_2}} \nonumber & := 3\,\ton{e^3\,\cos\ton{f - 2 \omega} + 6\,e\,\grf{\qua{2\,e\,+ \ton{4 + e^2}\,\cos f}\,\cos 2u +
e\,\cos\ton{4f + 2\omega}} + e^3\,\cos\ton{5 f + 2 \omega} - \right.\acap
&\left. - 16\,\sin f\,\sin\ton{f + 2 \omega}} - \qua{f \rightarrow f_0},\acap
\mathcal{A}_4^{\ton{J_2}}  & :=0,\acap
\mathcal{A}_5^{\ton{J_2}}  & :=0,\acap
\mathcal{A}_6^{\ton{J_2}} \nonumber & :=
6\,\ton{16\,\cos\ton{f + 2\omega}\,\sin f + e\,\grf{-e^2\,\sin\ton{f - 2\omega} + 6\,\qua{2\,e\,+ \ton{4 + e^2}\,\cos f}\,\sin 2u + \right.\right.\acap
&\left.\left. +  6\,e\,\sin\ton{4f + 2\omega} + e^2\,\sin\ton{5 f + 2\omega}}} - \qua{f \rightarrow f_0},\acap
\mathcal{E}_1^{\ton{J_2}} & := 4\,\qua{3\,\ton{4 + e^2}\,\cos f + e\,\ton{6\,\cos 2f + e\,\cos 3f}}  - \qua{f \rightarrow f_0},\acap
\mathcal{E}_2^{\ton{J_2}} & := -6\,\qua{3\,\ton{4 + e^2}\,\cos f + e\,\ton{6\,\cos 2f + e\,\cos 3f}}  - \qua{f \rightarrow f_0}, \acap
\mathcal{E}_3^{\ton{J_2}} \nonumber & := -4\,\qua{3\,\cos\ton{f + 2 \omega} + 7\,\cos\ton{3 f + 2 \omega}} +
e \grf{-e\,\qua{3\,\cos\ton{f - 2 \omega} + \right.\right.\acap
\nonumber &\left.\left. + 33\,\cos\ton{f + 2 \omega} + 17\,\cos\ton{3 f + 2 \omega} + 3\,\cos\ton{5 f + 2 \omega}} + 36\,\sin 2f\,\sin 2u +\right.\acap
&\left. + 120\,\sin f\,\sin\ton{f + 2 \omega}}  - \qua{f \rightarrow f_0}, \acap
\mathcal{E}_4^{\ton{J_2}} & := 0,\acap
\mathcal{E}_5^{\ton{J_2}} & := 0,\acap
\mathcal{E}_6^{\ton{J_2}} \nonumber & := 6\,e^2\,\sin\ton{f - 2 \omega} - 8\,\qua{3\,\sin\ton{f + 2 \omega} + 7\,\sin\ton{3 f + 2 \omega}} -
2\,e\,\grf{24\,\qua{3\,\cos f\,\cos 2u \right.\right.\acap
&\left.\left. +  5\,\cos\ton{f + 2 \omega}}\,\sin f + e\,\qua{33\,\sin\ton{f + 2 \omega} + 17\,\sin\ton{3 f + 2 \omega} + 3\,\sin\ton{5 f + 2 \omega}}}  - \qua{f \rightarrow f_0},\acap
\mathcal{I}_1^{\ton{J_2}}  & :=0,\acap
\mathcal{I}_2^{\ton{J_2}}  & :=0,\acap
\mathcal{I}_3^{\ton{J_2}}  & :=0,\acap
\mathcal{I}_4^{\ton{J_2}} & := 6 f + 6\,e\,\sin f + 3\,\sin 2u + 3\,e\,\sin\ton{f + 2 \omega} + e\,\sin\ton{3 f + 2 \omega}  - \qua{f \rightarrow f_0}, \acap
\mathcal{I}_5^{\ton{J_2}} & := -\grf{3\,\cos 2u + e\,\qua{3\,\cos\ton{f + 2 \omega} + \cos\ton{3 f + 2 \omega}}}  - \qua{f \rightarrow f_0}, \acap
\mathcal{I}_6^{\ton{J_2}}  & :=0,\acap
\mathcal{N}_1^{\ton{J_2}}  & :=0,\acap
\mathcal{N}_2^{\ton{J_2}}  & :=0,\acap
\mathcal{N}_3^{\ton{J_2}}  & :=0,\acap
\mathcal{N}_4^{\ton{J_2}} & := - \grf{3\,\cos 2u + e\,\qua{3\,\cos\ton{f + 2 \omega} + \cos\ton{3 f + 2 \omega}}}  - \qua{f \rightarrow f_0}, \acap
\mathcal{N}_5^{\ton{J_2}}  & := 6 f + 6\,e\,\sin f - 3\,\sin 2u - e\,\qua{3\,\sin\ton{f + 2 \omega} + \sin\ton{3 f + 2 \omega}}  - \qua{f \rightarrow f_0}, \acap
\mathcal{N}_6^{\ton{J_2}}  & :=0,\acap
\mathcal{P}_1^{\ton{J_2}} & := 48\,e\,f + 8\,\ton{6 + 5\,e^2 + 6\,e\,\cos f + e^2\,\cos 2f}\,\sin f - \qua{f \rightarrow f_0},\acap
\mathcal{P}_2^{\ton{J_2}} & := 6\,\qua{-12\,e\,f - 2\,\ton{6 + 5\,e^2 + 6\,e\,\cos f + e^2\,\cos 2f}\,\sin f} - \qua{f \rightarrow f_0}, \acap
\mathcal{P}_3^{\ton{J_2}}  \nonumber & := 4\,\qua{3\,\sin\ton{f + 2 \omega} - 7\,\sin\ton{3 f + 2 \omega}} -
e \grf{36\,\qua{3\,\cos\ton{f + 2 \omega} + \cos\ton{3 f + 2 \omega}}\,\sin f + \right.\acap
&\left. + e\,\qua{3\,\sin\ton{f - 2 \omega} + 21\,\sin\ton{f + 2 \omega} +
11\,\sin\ton{3 f + 2 \omega} + 3\,\sin\ton{5 f + 2 \omega}}} - \qua{f \rightarrow f_0}, \acap
\mathcal{P}_4^{\ton{J_2}}  & := -8\,e\,\grf{3\,\cos 2u + e\,\qua{3\,\cos\ton{f + 2 \omega} + \cos\ton{3 f + 2 \omega}}}  \cot I
- \qua{f \rightarrow f_0}, \acap
\mathcal{P}_5^{\ton{J_2}}  & := -8\,e\, \cot I \grf{-6 f + 3\,\sin 2u + e\,\qua{-6\,\sin f + 3\,\sin\ton{f + 2 \omega} + \sin\ton{3 f + 2 \omega}}}
- \qua{f \rightarrow f_0}, \acap
\mathcal{P}_6^{\ton{J_2}}  \nonumber & := -6\,e^2\,\cos\ton{f - 2 \omega} + 6 (-4 + 7\,e^2)\,\cos\ton{f + 2 \omega} +
56\,\cos\ton{3 f + 2 \omega} + 2\,e\,\grf{11\,e\,\cos\ton{3 f + 2 \omega} + \right.\acap
&\left. + 3\,e\,\cos\ton{5 f + 2 \omega} - 36\,\sin f\,\qua{3\,\sin\ton{f + 2 \omega} + \sin\ton{3 f + 2 \omega}}} - \qua{f \rightarrow f_0}.
\end{align}
}

\subsection{\textcolor{black}{Coefficients of the total mixed shifts per orbit of order $J_2/c^2$}}\lb{appenA2}
\renewcommand{\theequation}{A.2.\arabic{equation}}
\setcounter{equation}{0}
\textcolor{black}{
Here, the mixed averaged shifts per orbit of order $\mathcal{O}\ton{J_2/c^2}$, calculated in Section\,\ref{j2c2}, are treated.
The explicit expressions of the coefficients $\mathcal{\overline{A}}_1^{\ton{J_2/c^2}},\ldots \mathcal{\overline{H}}_6^{\ton{J_2/c^2}}$ entering \rfrs{coeff_sma}{coeff_eta} are displayed below.
They read
}
\textcolor{black}{
\begin{align}
\mathcal{\overline{A}}_1^{\ton{J_2/c^2}} & :=0,\acap
\mathcal{\overline{A}}_2^{\ton{J_2/c^2}} & :=0,\acap
\mathcal{\overline{A}}_3^{\ton{J_2/c^2}} \nonumber \lb{sma1}& := 8\,\ton{1 + e\,\cos f_0}^3\,\cos 2\omega\,\sin 2f_0 + \grf{4\,e\,\ton{3 + e^2}\,\cos f_0 + 4\,\ton{2 + 3\,e^2}\,\cos 2f_0 + e\,\qua{3\,\ton{4 + e^2}\,\cos 3f_0 + \right.\right.\acap
& \left.\left.  + e\,\ton{12 + e^2 + 6\,\cos 4f_0 + e\,\cos 5f_0}}}\,\sin 2\omega, \acap
\mathcal{\overline{A}}_4^{\ton{J_2/c^2}} & :=0,\acap
\mathcal{\overline{A}}_5^{\ton{J_2/c^2}} & :=0,\acap
\mathcal{\overline{A}}_6^{\ton{J_2/c^2}} \nonumber & := -2 \grf{4\,e\,\ton{3 + e^2}\,\cos f_0 + 4\,\ton{2 + 3\,e^2}\,\cos 2f_0 + e\,\qua{3\,\ton{4 + e^2}\,\cos 3f_0 + e\,\ton{12 + e^2 + 6\,\cos 4f_0 + e\,\cos 5f_0}}}\,\cos 2\omega + \acap
& + 16\,\ton{1 + e\,\cos f_0}^3\,\sin 2f_0\,\sin 2\omega, \acap
\mathcal{\overline{E}}_1^{\ton{J_2/c^2}} & :=0,\acap
\mathcal{\overline{E}}_2^{\ton{J_2/c^2}} & :=0,\acap
\mathcal{\overline{E}}_3^{\ton{J_2/c^2}} \nonumber & := - \grf{4\,\qua{3\,\sin\ton{f_0 + 2 \omega} + 7\,\sin\ton{3 f_0 + 2 \omega}} +
e\,\qua{-3\,e\,\sin\ton{f_0 - 2 \omega} + \ton{20 + 19\,e^2}\,\sin 2\omega + 60\,\sin u_0 + 18\,\sin\ton{4f_0 + 2\omega} +\right.\right.\acap
&\left.\left. + 33\,e\,\sin\ton{f_0 + 2 \omega} + 17\,e\,\sin\ton{3 f_0 + 2 \omega} + 3\,e\,\sin\ton{5 f_0 + 2 \omega}}}, \acap
\mathcal{\overline{E}}_4^{\ton{J_2/c^2}} & :=0,\acap
\mathcal{\overline{E}}_5^{\ton{J_2/c^2}} & :=0,\acap
\mathcal{\overline{E}}_6^{\ton{J_2/c^2}} \nonumber & := 8\,\qua{3\,\cos\ton{f_0 + 2 \omega} + 7\,\cos\ton{3 f_0 + 2 \omega}} + 2\,e\,\qua{3\,e\,\cos\ton{f_0 - 2 \omega} + \ton{20 + 19\,e^2}\,\cos 2\omega + 60\,\cos u_0 + 18\,\cos\ton{4f_0 + 2\omega} + \right.\acap
&\left. + 33\,e\,\cos\ton{f_0 + 2 \omega} + 17\,e\,\cos\ton{3 f_0 + 2 \omega} + 3\,e\,\cos\ton{5 f_0 + 2 \omega}}, \acap
\mathcal{\overline{I}}_1^{\ton{J_2/c^2}} & :=0,\acap
\mathcal{\overline{I}}_2^{\ton{J_2/c^2}} & :=0,\acap
\mathcal{\overline{I}}_3^{\ton{J_2/c^2}} & :=0,\acap
\mathcal{\overline{I}}_4^{\ton{J_2/c^2}} & := 5\,e^2 + 3\,\cos u_0 + e\,\qua{-16\,\cos f_0 + 2\,e\,\cos 2\omega + 3\,\cos\ton{f_0 + 2 \omega} + \cos\ton{3 f_0 + 2 \omega}}, \acap
\mathcal{\overline{I}}_5^{\ton{J_2/c^2}} & := 3\,\sin u_0 + e\,\qua{2\,e\,\sin 2\omega + 3\,\sin\ton{f_0 + 2 \omega} + \sin\ton{3 f_0 + 2 \omega}}, \acap
\mathcal{\overline{I}}_6^{\ton{J_2/c^2}} & :=0,\acap
\mathcal{\overline{N}}_1^{\ton{J_2/c^2}} & :=0,\acap
\mathcal{\overline{N}}_2^{\ton{J_2/c^2}} & :=0,\acap
\mathcal{\overline{N}}_3^{\ton{J_2/c^2}} & :=0,\acap
\mathcal{\overline{N}}_4^{\ton{J_2/c^2}} & := 3\,\sin u_0 + e\,\qua{2\,e\,\sin 2\omega + 3\,\sin\ton{f_0 + 2 \omega} + \sin\ton{3 f_0 + 2 \omega}}, \acap
\mathcal{\overline{N}}_5^{\ton{J_2/c^2}} & := 5\,e^2 - 3\,\cos u_0 - e\,\qua{16\,\cos f_0 + 2\,e\,\cos 2\omega + 3\,\cos\ton{f_0 + 2 \omega} + \cos\ton{3 f_0 + 2 \omega}}, \acap
\mathcal{\overline{N}}_6^{\ton{J_2/c^2}} & :=0,\acap
\mathcal{\overline{P}}_1^{\ton{J_2/c^2}} & := -4\,e\,\ton{44 + 17\,e^2 - 64\,e\,\cos f_0}\,\sin I, \acap
\mathcal{\overline{P}}_2^{\ton{J_2/c^2}} & := 6\,e\,\ton{44 + 17\,e^2 - 64\,e\,\cos f_0}\,\sin I, \acap
\mathcal{\overline{P}}_3^{\ton{J_2/c^2}} \nonumber & := 2 \grf{4\,\qua{-3\,\cos\ton{f_0 + 2 \omega} + 7\,\cos\ton{3 f_0 + 2 \omega}} + e\,\qua{-3\,e\,\cos\ton{f_0 - 2 \omega} + 2\,\ton{-10 + 9\,e^2}\,\cos 2\omega + 60\,\cos u_0 + \right.\right.\acap
&\left.\left. + 18\,\cos\ton{4f_0 + 2\omega} + 45\,e\,\cos\ton{f_0 + 2 \omega} + 19\,e\,\cos\ton{3 f_0 + 2 \omega} + 3\,e\,\cos\ton{5 f_0 + 2 \omega}}}\,\sin I, \acap
\mathcal{\overline{P}}_4^{\ton{J_2/c^2}} & := -16\,e\,\cos I \grf{3\,\sin u_0 + e\,\qua{2\,e\,\sin 2\omega + 3\,\sin\ton{f_0 + 2 \omega} + \sin\ton{3 f_0 + 2 \omega}}},\acap
\mathcal{\overline{P}}_5^{\ton{J_2/c^2}} & := 16\,e\,\cos I \grf{-5\,e^2 + 3\,\cos u_0 + e\,\qua{16\,\cos f_0 + 2\,e\,\cos 2\omega + 3\,\cos\ton{f_0 + 2 \omega} +
\cos\ton{3 f_0 + 2 \omega}}},\acap
\mathcal{\overline{P}}_6^{\ton{J_2/c^2}} \nonumber & := 4\,\sin I \grf{4\,\qua{-3\,\sin\ton{f_0 + 2 \omega} + 7\,\sin\ton{3 f_0 + 2 \omega}} + e\,\qua{3\,e\,\sin\ton{f_0 - 2 \omega} + 2\,\ton{-10 + 9\,e^2}\,\sin 2\omega + 60\,\sin u_0 + \right.\right.\acap
&\left.\left. + 18\,\sin\ton{4f_0 + 2\omega} + 45\,e\,\sin\ton{f_0 + 2 \omega} + 19\,e\,\sin\ton{3 f_0 + 2 \omega} + 3\,e\,\sin\ton{5 f_0 + 2 \omega}}},\acap
\mathcal{\overline{H}}_1^{\ton{J_2/c^2}} \nonumber & := 4\,e\,\grf{88 + 5\,e^4 - 16\,\sqrt{1 - e^2} - 3\,e^2\,\ton{21 + 8\,\sqrt{1 - e^2}} - e\,\qua{3\,e^2\,\ton{7 + 4\,\sqrt{1 - e^2}} + 8\,\ton{-17 + 6\,\sqrt{1 - e^2}}}\,\cos f_0 + \right.\acap
&\left. + e^2\,\qua{8\,\ton{5 - 3\,\sqrt{1 - e^2}}\,\cos 2f_0 + e\,\ton{5 - 4\,\sqrt{1 - e^2}}\,\cos 3f_0}}, \acap
\mathcal{\overline{H}}_2^{\ton{J_2/c^2}} \nonumber & := 6\,e\,\grf{-88 - 5\,e^4 + 16\,\sqrt{1 - e^2} + 3\,e^2\,\ton{21 + 8\,\sqrt{1 - e^2}} + e\,\qua{3\,e^2\,\ton{7 + 4\,\sqrt{1 - e^2}} + 8\,\ton{-17 + 6\,\sqrt{1 - e^2}}}\,\cos f_0 + \right.\acap
&\left. + 4\,e^2\,\sqrt{1 - e^2}\,\ton{6\,\cos 2f_0 + e\,\cos 3f_0} - 5\,e^2\,\ton{8\,\cos 2f_0 + e\,\cos 3f_0}}, \acap
\mathcal{\overline{H}}_3^{\ton{J_2/c^2}} \nonumber & := 3\,e^2\,\ton{2 - 7\,e^2}\,\cos\ton{f_0 - 2 \omega} + 96\,e\,\sqrt{1 - e^2}\,\ton{1 + e\,\cos f_0}^3\,\cos u_0 +
8\,\qua{3\,\cos\ton{f_0 + 2 \omega} - 7\,\cos\ton{3 f_0 + 2 \omega}} + \acap
\beg + e\,\qua{-2\,\ton{-20 + 7\,e^2 + 13\,e^4}\,\cos 2\omega - 12\,\ton{14 + 11\,e^2}\,\cos u_0 - 18\,\ton{2 + 3\,e^2}\,\cos\ton{4f_0 + 2\omega} - \right.\acap
&\left. - 3\,e\,\ton{74 + 9\,e^2}\,\cos\ton{f_0 + 2 \omega} - e\,\ton{138 + 31\,e^2}\,\cos\ton{3 f_0 + 2 \omega} - 3\,e\,\ton{2 + 3\,e^2}\,\cos\ton{5 f_0 + 2 \omega}}, \acap
\mathcal{\overline{H}}_4^{\ton{J_2/c^2}} & :=0, \acap
\mathcal{\overline{H}}_5^{\ton{J_2/c^2}} & :=0, \acap
\mathcal{\overline{H}}_6^{\ton{J_2/c^2}} \nonumber \lb{eta6}& := -2\,\qua{3\,e^2\,\qua{2 + e^2\,\ton{-7 + 4\,\sqrt{1 - e^2}}}\,\sin\ton{f_0 - 2 \omega} +
2\,e\,\qua{-20 + 13\,e^4 + e^2\,\ton{7 - 36\,\sqrt{1 - e^2}}}\,\sin 2\omega + \right.\acap
\nonumber &\left. + 8\,\qua{-3\,\sin\ton{f_0 + 2 \omega} + 7\,\sin\ton{3 f_0 + 2 \omega}} +
e\,\ton{12\,\qua{14 - 8\,\sqrt{1 - e^2} + e^2\,\ton{11 - 12\,\sqrt{1 - e^2}}}\,\sin u_0 + \right.\right.\acap
\nonumber &\left.\left. + 18\,\qua{2 + e^2\,\ton{3 - 4\,\sqrt{1 - e^2}}}\,\sin\ton{4f_0 + 2\omega} +
e \grf{3\,\qua{74 - 48\,\sqrt{1 - e^2} + e^2\,\ton{9 - 12\,\sqrt{1 - e^2}}}\,\sin\ton{f_0 + 2 \omega} + \right.\right.\right.\acap
&\left.\left.\left. + \qua{138 - 144\,\sqrt{1 - e^2} + e^2\,\ton{31 - 36\,\sqrt{1 - e^2}}}\,\sin\ton{3 f_0 + 2 \omega} + 3\,\qua{2 + e^2\,\ton{3 - 4\,\sqrt{1 - e^2}}}\,\sin\ton{5 f_0 + 2 \omega}}}}.
\end{align}
}
\end{appendices}
\section*{Conflict of interest statement}
I declare no conflicts of interest.
\section*{Data availability}
No new data were generated or analysed in support of this research.
\bibliography{Uranusbib}{}

\begin{thebibliography}{28}
\providecommand{\natexlab}[1]{#1}
\providecommand{\url}[1]{\texttt{#1}}
\expandafter\ifx\csname urlstyle\endcsname\relax
  \providecommand{\doi}[1]{doi: #1}\else
  \providecommand{\doi}{doi: \begingroup \urlstyle{rm}\Url}\fi

\bibitem[{Bertotti} et~al.(2003){Bertotti}, {Farinella}, and
  {Vokrouhlick\'{y}}]{2003ASSL..293.....B}
B.~{Bertotti}, P.~{Farinella}, and D.~{Vokrouhlick\'{y}}.
\newblock \emph{{Physics of the Solar System}}.
\newblock Kluwer, Dordrecht, Aug 2003.
\newblock \doi{10.1007/978-94-010-0233-2}.

\bibitem[{Bolton} et~al.(2017){Bolton}, {Lunine}, {Stevenson}, {Connerney},
  {Levin}, {Owen}, {Bagenal}, {Gautier}, {Ingersoll}, {Orton}, {Guillot},
  {Hubbard}, {Bloxham}, {Coradini}, {Stephens}, {Mokashi}, {Thorne}, and
  {Thorpe}]{2017SSRv..213....5B}
S.~J. {Bolton}, J.~{Lunine}, D.~{Stevenson}, J.~E.~P. {Connerney}, S.~{Levin},
  T.~C. {Owen}, F.~{Bagenal}, D.~{Gautier}, A.~P. {Ingersoll}, G.~S. {Orton},
  T.~{Guillot}, W.~{Hubbard}, J.~{Bloxham}, A.~{Coradini}, S.~K. {Stephens},
  P.~{Mokashi}, R.~{Thorne}, and R.~{Thorpe}.
\newblock {The Juno Mission}.
\newblock \emph{Space Sci. Rev.}, 213\penalty0 (1-4):\penalty0 5--37, Nov 2017.
\newblock \doi{10.1007/s11214-017-0429-6}.

\bibitem[{Brumberg}(1991)]{1991ercm.book.....B}
V.~A. {Brumberg}.
\newblock \emph{{Essential Relativistic Celestial Mechanics}}.
\newblock Adam Hilger, Bristol, 1991.

\bibitem[{Capderou}(2005)]{Capde05}
M.~{Capderou}.
\newblock \emph{Satellites. Orbits and Missions}.
\newblock Springer-Verlag France, Paris, Mar 2005.
\newblock \doi{10.1007/b139118}.

\bibitem[{Ciufolini} et~al.(2013){Ciufolini}, {Paolozzi}, {Koenig}, {Pavlis},
  {Ries}, {Matzner}, {Gurzadyan}, {Penrose}, {Sindoni}, and
  {Paris}]{2013NuPhS.243..180C}
I.~{Ciufolini}, A.~{Paolozzi}, R.~{Koenig}, E.~C. {Pavlis}, J.~{Ries},
  R.~{Matzner}, V.~{Gurzadyan}, R.~{Penrose}, G.~{Sindoni}, and C.~{Paris}.
\newblock {Fundamental Physics and General Relativity with the LARES and LAGEOS
  satellites}.
\newblock \emph{Nucl. Phys. B Proc. Suppl.}, 243:\penalty0 180--193, Oct 2013.
\newblock \doi{10.1016/j.nuclphysbps.2013.09.005}.

\bibitem[{Debono} and {Smoot}(2016)]{2016Univ....2...23D}
I.~{Debono} and G.~F. {Smoot}.
\newblock {General Relativity and Cosmology: Unsolved Questions and Future
  Directions}.
\newblock \emph{Universe}, 2:\penalty0 23, Sep 2016.
\newblock \doi{10.3390/universe2040023}.

\bibitem[{Einstein}(1915)]{Ein15}
A.~{Einstein}.
\newblock {Erkl\"{a}rung der Perihelbewegung des Merkur aus der allgemeinen
  Relativit\"{a}tstheorie}.
\newblock \emph{Sitzber. Preuss. Akad.}, 47:\penalty0 831--839, Nov 1915.

\bibitem[{Gurfil} and {Efroimsky}(2022)]{2022AdSpR..69..538G}
P.~{Gurfil} and M.~{Efroimsky}.
\newblock {Analysis of the PPN two-Body Problem using non-osculating orbital
  elements}.
\newblock \emph{Adv. Space Res.}, 69\penalty0 (1):\penalty0 538--553, Jan 2022.
\newblock \doi{10.1016/j.asr.2021.09.009}.

\bibitem[{Heimberger} et~al.(1989){Heimberger}, {Soffel}, and
  {Ruder}]{1990CeMDA..47..205H}
J.~{Heimberger}, M.~{Soffel}, and H.~{Ruder}.
\newblock {Relativistic effects in the motion of artificial satellites - The
  oblateness of the central body II}.
\newblock \emph{Celest. Mech. Dyn. Astr.}, 47\penalty0 (2):\penalty0 205--217,
  Jun 1989.
\newblock \doi{10.1007/BF00051205}.

\bibitem[{Huang} and {Liu}(1992)]{1992CeMDA..53..293H}
C.~{Huang} and L.~{Liu}.
\newblock {Analytical solutions to the four post-Newtonian effects in a near
  Earth satellite orbit}.
\newblock \emph{Celest. Mech. Dyn. Astr.}, 53\penalty0 (3):\penalty0 293--307,
  Sep 1992.
\newblock \doi{10.1007/BF00052615}.

\bibitem[{Iorio}(2015)]{2015IJMPD..2450067I}
L.~{Iorio}.
\newblock {Post-Newtonian direct and mixed orbital effects due to the
  oblateness of the central body}.
\newblock \emph{Int. J. Mod. Phys. D}, 24\penalty0 (8):\penalty0 1550067-59,
  Jun 2015.
\newblock \doi{10.1142/S0218271815500674}.

\bibitem[{Iorio}(2016)]{2016MNRAS.460.2445I}
L.~{Iorio}.
\newblock {Post-Keplerian corrections to the orbital periods of a two-body
  system and their measurability}.
\newblock \emph{Mon. Not. Roy. Astron. Soc.}, 460\penalty0 (3):\penalty0
  2445--2452, May 2016.
\newblock \doi{10.1093/mnras/stw1155}.

\bibitem[{Iorio}(2019{\natexlab{a}})]{2019AJ....157..220I}
L.~{Iorio}.
\newblock {Calculation of the Uncertainties in the Planetary Precessions with
  the Recent EPM2017 Ephemerides and their Use in Fundamental Physics and
  Beyond}.
\newblock \emph{Astron J.}, 157\penalty0 (6):\penalty0 220, Jun
  2019{\natexlab{a}}.
\newblock \doi{10.3847/1538-3881/ab19bf}.

\bibitem[{Iorio}(2019{\natexlab{b}})]{2019EPJC...79..816I}
L.~{Iorio}.
\newblock {On the mean anomaly and the mean longitude in tests of
  post-Newtonian gravity}.
\newblock \emph{Eur. Phys. J. C}, 79\penalty0 (10):\penalty0 816, Oct
  2019{\natexlab{b}}.
\newblock \doi{10.1140/epjc/s10052-019-7337-8}.

\bibitem[{Iorio}(2023)]{2023arXiv231002834I}
L.~{Iorio}.
\newblock {Post-Newtonian orbital effects induced by the mass quadrupole and
  spin octupole moments of an axisymmetric body}.
\newblock \emph{arXiv e-prints}, art. arXiv:2310.02834, Oct 2023.
\newblock \doi{10.48550/arXiv.2310.02834}.

\bibitem[{King-Hele}(1958)]{1958RSPSA.247...49K}
D.~G. {King-Hele}.
\newblock {The Effect of the Earth's Oblateness on the Orbit of a Near
  Satellite}.
\newblock \emph{Proc. R. Soc. A}, 247\penalty0 (1248):\penalty0 49--72, Sep
  1958.
\newblock \doi{10.1098/rspa.1958.0169}.

\bibitem[{Klioner} and {Kopeikin}(1994)]{1994ApJ...427..951K}
S.~A. {Klioner} and S.~M. {Kopeikin}.
\newblock {The Post-Keplerian Orbital Representations of the Relativistic
  Two-Body Problem}.
\newblock \emph{Astrophys. J}, 427:\penalty0 951, Jun 1994.
\newblock \doi{10.1086/174201}.

\bibitem[{Kopeikin} et~al.(2011){Kopeikin}, {Efroimsky}, and
  {Kaplan}]{2011rcms.book.....K}
S.~M. {Kopeikin}, M.~{Efroimsky}, and G.~{Kaplan}.
\newblock \emph{{Relativistic Celestial Mechanics of the Solar System}}.
\newblock Wiley-VCH, Weinheim, Aug 2011.
\newblock \doi{10.1002/9783527634569}.

\bibitem[{Le Verrier}(1859)]{LeVer1859}
U.~{Le Verrier}.
\newblock {Lettre de M. Le Verrier \`{a} M. Faye sur la th\'{e}orie de Mercure
  et sur le mouvement du p\'{e}rih\'{e}lie de cette plan\`{e}te}.
\newblock \emph{Cr. Hebd. Acad. Sci.}, 49:\penalty0 379--383, Aug 1859.

\bibitem[{Nobili} and {Will}(1986)]{1986Natur.320...39N}
A.~M. {Nobili} and C.~M. {Will}.
\newblock {The real value of Mercury's perihelion advance}.
\newblock \emph{Nature}, 320:\penalty0 39--41, Mar 1986.
\newblock \doi{10.1038/320039a0}.

\bibitem[{Pearlman} et~al.(2019){Pearlman}, {Arnold}, {Davis}, {Barlier},
  {Biancale}, {Vasiliev}, {Ciufolini}, {Paolozzi}, {Pavlis}, {So{\'s}nica}, and
  {Blo{\ss}feld}]{2019JGeod..93.2181P}
M.~{Pearlman}, D.~{Arnold}, M.~{Davis}, F.~{Barlier}, R.~{Biancale},
  V.~{Vasiliev}, I.~{Ciufolini}, A.~{Paolozzi}, E.~C. {Pavlis},
  K.~{So{\'s}nica}, and M.~{Blo{\ss}feld}.
\newblock {Laser geodetic satellites: a high-accuracy scientific tool}.
\newblock \emph{J. Geod.}, 93\penalty0 (11):\penalty0 2181--2194, Nov 2019.
\newblock \doi{10.1007/s00190-019-01228-y}.

\bibitem[{Pitjeva} and {Pitjev}(2018)]{2018AstL...44..554P}
E.~V. {Pitjeva} and N.~P. {Pitjev}.
\newblock {Masses of the Main Asteroid Belt and the Kuiper Belt from the
  Motions of Planets and Spacecraft}.
\newblock \emph{Astron. Lett.}, 44\penalty0 (8-9):\penalty0 554--566, Aug 2018.
\newblock \doi{10.1134/S1063773718090050}.

\bibitem[{Poisson} and {Will}(2014)]{2014grav.book.....P}
E.~{Poisson} and C.~M. {Will}.
\newblock \emph{{Gravity}}.
\newblock Cambridge University Press, Cambridge, Jun 2014.
\newblock \doi{10.1017/CBO9781139507486}.

\bibitem[{Renzetti}(2013)]{2013CEJPh..11..531R}
G.~{Renzetti}.
\newblock {History of the attempts to measure orbital frame--dragging with
  artificial satellites}.
\newblock \emph{Centr. Eur. J. Phys.}, 11\penalty0 (5):\penalty0 531--544, Jul
  2013.
\newblock \doi{10.2478/s11534-013-0189-1}.

\bibitem[{Soffel}(1989)]{Sof89}
M.~H. {Soffel}.
\newblock \emph{Relativity in Astrometry, Celestial Mechanics and Geodesy}.
\newblock Springer, Heidelberg, 1989.
\newblock \doi{10.1007/978-3-642-73406-9}.

\bibitem[{Soffel} and {Han}(2019)]{SoffelHan19}
M.~H. {Soffel} and W.-B. {Han}.
\newblock \emph{{Applied General Relativity}}.
\newblock {Astronomy and Astrophysics Library}. Springer Nature Switzerland,
  Cham, Oct 2019.
\newblock \doi{10.1007/978-3-030-19673-8}.

\bibitem[{Soffel} et~al.(1987){Soffel}, {Wirrer}, {Schastok}, {Ruder}, and
  {Schneider}]{1988CeMec..42...81S}
M.~H. {Soffel}, R.~{Wirrer}, J.~{Schastok}, H.~{Ruder}, and M.~{Schneider}.
\newblock {Relativistic Effects in the Motion of Artificial Satellites - the
  Oblateness of the Central Body I}.
\newblock \emph{Celest. Mech. Dyn. Astr.}, 42\penalty0 (1-4):\penalty0 81--89,
  Mar 1987.
\newblock \doi{10.1007/BF01232949}.

\bibitem[{Will}(2014)]{2014PhRvD..89d4043W}
C.~M. {Will}.
\newblock {Incorporating post-Newtonian effects in N-body dynamics}.
\newblock \emph{Phys. Rev. D}, 89\penalty0 (4):\penalty0 044043, Feb 2014.
\newblock \doi{10.1103/PhysRevD.89.044043}.

\end{thebibliography}
\end{document}